\documentclass[useAMS,usenatbib]{mnras}
\usepackage{aas_macros}
\usepackage{graphicx}
\usepackage{lscape}
\usepackage{mathptmx}
\usepackage{times}
\usepackage{amssymb}
\usepackage{hyperref}
\usepackage{siunitx}
\usepackage{calrsfs}
\usepackage[T1]{fontenc}
\usepackage{aecompl}
\DeclareMathAlphabet{\pazocal}{OMS}{zplm}{m}{n}
 
\def\p0{\phantom{0}}

\newcommand{\affil}[1]{$^{\rm #1}$}
\newcounter{inst} 
\newcommand{\inst}[1]{\noindent%
   \refstepcounter{inst}\affil{\arabic{inst}\label{#1}}     
   }

\begin{document}  
\title[High-resolution Observations of Low-luminosity GPS and CSS sources]{High-resolution Observations of Low-luminosity Gigahertz-Peaked Spectrum and Compact Steep Spectrum Sources}

\author[J.~D.~Collier et al. 2018]
{J.~D.~Collier\affil{\ref{CASS},\ref{WSU}}\thanks{E-mail: \href{mailto:Jordan.Collier@csiro.au}{Jordan.Collier@csiro.au}}, 
S.~J.~Tingay\affil{\ref{ICRAR},\ref{CAASTRO}}, 
J.~R.~Callingham\affil{\ref{CASS},\ref{CAASTRO},\ref{USyd}}, 
R.~P.~Norris\affil{\ref{CASS},\ref{WSU},\ref{CAASTRO}}, 
M.~D.~Filipovi\'c\affil{\ref{WSU}}, 
\newauthor 
T.~J.~Galvin\affil{\ref{CASS},\ref{WSU}, \ref{ICRAR}}, 
M.~T.~Huynh\affil{\ref{CASS_WA},\ref{UWA}} 
H.~T.~Intema\affil{\ref{NRAO},\ref{Leiden}}, 
J.~Marvil\affil{\ref{CASS}}, 
A.~N.~O'Brien\affil{\ref{CASS},\ref{WSU}}, 
Q.~Roper\affil{\ref{WSU}}, 
\newauthor
S.~Sirothia\affil{\ref{SKA_RSA},\ref{Rhodes}}, 
N.~F.~H.~Tothill\affil{\ref{WSU}}, 
M.~E.~Bell\affil{\ref{CASS},\ref{CAASTRO},\ref{UTS}}, 
B.-Q.~For\affil{\ref{UWA}}, 
B.~M.~Gaensler\affil{\ref{CAASTRO},\ref{USyd},\ref{Toronto}}, 
\newauthor
P.~J.~Hancock\affil{\ref{ICRAR},\ref{CAASTRO}}, 
L.~Hindson\affil{\ref{VUW},\ref{Hert}}, 
N.~Hurley-Walker\affil{\ref{ICRAR}}, 
M.~Johnston-Hollitt\affil{\ref{VUW},\ref{Peripety}},  
\newauthor
A.~D.~Kapi\'nska\affil{\ref{CAASTRO},\ref{UWA}}, 
E.~Lenc\affil{\ref{CAASTRO},\ref{USyd}}, 
J.~Morgan\affil{\ref{ICRAR}}, 
P.~Procopio\affil{\ref{UMelb}}, 
L.~Staveley-Smith\affil{\ref{CAASTRO},\ref{UWA}},  
\newauthor  
R.~B.~Wayth\affil{\ref{ICRAR},\ref{CAASTRO}}, 
C.~Wu\affil{\ref{UWA}}, 
Q.~Zheng\affil{\ref{VUW},\ref{Peripety}}, 
I.~Heywood\affil{\ref{CASS},\ref{Rhodes},\ref{Oxford}},
and 
A.~Popping\affil{\ref{ICRAR},\ref{CAASTRO}}\\
{\small \inst{CASS}\,CSIRO Astronomy and Space Science (CASS), Marsfield, NSW 2122, Australia}\\ 
{\small \inst{WSU}\,Western Sydney University, Locked Bag 1797, Penrith, NSW 2751, Australia}\\
{\small \inst{ICRAR}\,International Centre for Radio Astronomy Research (ICRAR), Curtin University, Bentley, WA 6102, Australia}\\ 
{\small \inst{CAASTRO}\,ARC Centre of Excellence for All-Sky Astrophysics (CAASTRO), Australia}\\
{\small \inst{USyd}\,Sydney Institute for Astronomy (SIfA), School of Physics, The University of Sydney, NSW 2006, Australia}\\
{\small \inst{CASS_WA}\,CSIRO Astronomy and Space Science (CASS), 26 Dick Perry Avenue, Kensington, WA 6151, Australia}\\
{\small \inst{UWA}\,International Centre for Radio Astronomy Research (ICRAR), M468, University of Western Australia, Crawley, WA 6009, Australia}\\
{\small \inst{NRAO}\,National Radio Astronomy Observatory (NRAO), 1003 Lopezville Road, Socorro, NM 87801-0387, USA}\\
{\small \inst{Leiden}\,Leiden Observatory, Leiden University, Niels Bohrweg 2, NL-2333CA, Leiden, The Netherlands}\\
{\small \inst{SKA_RSA}\,Square Kilometre Array South Africa, 3rd Floor, The Park, Park Road, Pinelands, 7405, South Africa}\\
{\small \inst{Rhodes}\,Department of Physics and Electronics, Rhodes University, PO Box 94, Grahamstown, 6140, South Africa}\\
{\small \inst{UTS}\,University of Technology Sydney, 15 Broadway, Ultimo NSW 2007, Australia}\\
{\small \inst{Toronto}\,Dunlap Institute for Astronomy and Astrophysics, 50 St. George St, University of Toronto, ON M5S 3H4, Canada}\\
{\small \inst{VUW}\,School of Chemical \& Physical Sciences, Victoria University of Wellington, PO Box 600, Wellington 6140, New Zealand}\\
{\small \inst{Hert}\, Centre for Astrophysics Research, School of Physics, Astronomy and Mathematics, University of Hertfordshire, College Lane, Hatfield AL10 9AB, UK}\\
{\small \inst{Peripety}\,Peripety Scientific Ltd., PO Box 11355 Manners Street, Wellington 6142, New Zealand}\\
{\small \inst{UMelb}\,School of Physics, The University of Melbourne, Parkville, VIC 3010, Australia}\\
{\small \inst{Oxford}\,Astrophysics, Department of Physics, University of Oxford, Keble Road, Oxford OX1 3RH, UK}
}

\date{$^\star$ E-mail: \href{mailto:Jordan.Collier@csiro.au}{Jordan.Collier@csiro.au}\\
Accepted 2018 February 23. Received 2018 February 23; in original form 2017 June 16}

\pagerange{\pageref{firstpage}--\pageref{lastpage}} \pubyear{2018}

\maketitle

\label{firstpage}

\begin{abstract}

We present Very Long Baseline Interferometry observations of a faint and low-luminosity ($L_{\rm 1.4 GHz} < 10^{27}$ W Hz$^{-1}$) Gigahertz-Peaked Spectrum (GPS) and Compact Steep Spectrum (CSS) sample. We select eight sources from deep radio observations that have radio spectra characteristic of a GPS or CSS source and an angular size of $\theta \lesssim 2\arcsec$, and detect six of them with the Australian Long Baseline Array. We determine their linear sizes, and model their radio spectra using Synchrotron Self Absorption (SSA) and Free Free Absorption (FFA) models. 
We derive statistical model ages, based on a fitted scaling relation, and spectral ages, based on the radio spectrum, which are generally consistent with the hypothesis that GPS and CSS sources are young and evolving. We resolve the morphology of one CSS source with a radio luminosity of $10^{25}~\mbox{W\,Hz}^{-1}$, and find what appear to be two hotspots spanning 1.7 kpc.
We find that our sources follow the turnover-linear size relation, and that both homogenous SSA and an inhomogeneous FFA model can account for the spectra with observable turnovers. All but one of the FFA models do not require a spectral break to account for the radio spectrum, while all but one of the alternative SSA and power law models do require a spectral break to account for the radio spectrum. We conclude that our low-luminosity sample is similar to brighter samples in terms of their spectral shape, turnover frequencies, linear sizes, and ages, but cannot test for a difference in morphology.

\end{abstract}


\section{Introduction}

Gigahertz-Peaked Spectrum (GPS) and Compact Steep Spectrum (CSS) sources are small but powerful Active Galactic Nuclei (AGN) that have a peaked radio spectrum with a characteristic turnover frequency, and are generally hosted by elliptical galaxies \citep{1997A&A...321..105D,2003PASA...20..118S,1994ApJS...91..491G,Orienti}. GPS sources turn over (i.e. reach a maximum radio flux density) at a few GHz, or above a few GHz in the sub-class of high frequency peakers (HFPs) defined by \cite{2000A&A...363..887D}. CSS sources turn over at a few hundred MHz or lower and have steep ($\alpha < -0.7$)\footnote{Throughout this paper, we use $S \propto \nu^{\alpha}$} spectral indices across the GHz range. Most GPS and CSS sources are symmetric, in which a two-sided structure is observed that resembles a scaled-down Fanaroff-Riley Type II (FR II) galaxy \citep{1974MNRAS.167P..31F}, consisting of steep-spectrum mini-lobes and hotspots, sometimes with a weak inverted or flat-spectrum core, and weak jets. Based on their linear size ($l$), GPS and CSS sources are morphologically classified as Compact Symmetric Objects (CSOs), which have linear sizes $l < 1$ kpc, or Medium-Sized Symmetric Objects \citep[MSOs;][]{1995A&A...302..317F,An12}, which have $l > 1$ kpc. 

It is widely accepted that GPS and CSS sources are young and evolving radio sources that may develop into large-scale radio sources \citep[e.g.][]{1995A&A...302..317F,ODea98,2000MNRAS.319....8A,2000MNRAS.319..445S,2003PASA...20...69P, 2006A&A...445..889T, 2009AN....330..120F,2011MNRAS.416.1135R,Orienti}. Evidence for this {\it youth} hypothesis includes their appearance as scaled-down versions of FR II galaxies, kinematic age estimates via proper motion measurements of their hotspot expansion speeds \citep{2009AN....330..193G,2003PASA...20...69P,2009AN....330..149P} and models of their radio spectra and spectral ages \citep{1999A&A...345..769M,2003PASA...20...19M}. If GPS/CSS sources are the youngest radio galaxies, then they are ideal objects for investigating the birth and early lives of radio emission in AGN.

However, the hypothesis that all GPS and CSS grow to large sizes is disputed, since statistical studies of the luminosity functions have revealed an over-abundance of the most compact sources relative to the number of large-scale radio galaxies \citep{1996ApJ...460..634R,1997AJ....113..148O,An12,Cal15}. 

The alternative {\it frustration} hypothesis is that GPS and CSS sources are frustrated by interactions with dense gas and dust in their environment, which halts the expansion of the jets \citep{1984AJ.....89....5V,1990A&A...232...19B}. Furthermore, some GPS and CSS sources have been interpreted as prematurely dying radio sources that switch off before growing to large sizes \citep{2009AN....330..303F,2010MNRAS.402.1892O} or recurrent radio galaxies \citep{1990A&A...232...19B,2012A&A...545A..91S}. The evolutionary model presented by \cite{An12} suggests that each of these scenarios exists amongst the GPS and CSS population, with only $\sim$30 per cent of sources evolving into large scale radio galaxies.

Many multi-frequency observations do not support the frustration hypothesis, which show that the host galaxies contain gas similar to FR II hosts \citep{1995A&A...302..317F,2000A&A...358..499F,2005ApJ...632..110S,Orienti}. However, many observational studies of individual GPS sources suggest the presence of a dense medium that may cause significant frustration \citep[e.g.][]{2014ApJ...780..178M,Cal15}. It is likely that amongst the GPS and CSS population as a whole is made up of both young and frustrated sources, individually or simultaneously. However, the significance of each contribution to the GPS and CSS population is generally unknown, especially at low luminosity. If frustration is minimal, youth is also likely to be present, but if frustration is dominant, youth is not necessary to explain their compactness. One reason these hypotheses are still debated is because the absorption mechanism responsible for the peaked spectra is still uncertain \citep{Cal15}.

\subsection{Radio spectra and absorption models}

\label{radio_sed_models}

The absorption mechanisms proposed to be responsible for the peaked spectra of GPS and CSS sources and the associated models \citep[see][]{Cal15} are summarised below.

\subsubsection{Synchrotron Self Absorption}

\label{sec_SSA}

The turnover in the spectra of GPS and CSS sources has generally been attributed to Synchrotron Self Absorption (SSA), related to their small size \citep{2000MNRAS.319..445S,2008A&A...487..885O,2009AN....330..120F,Orienti}. SSA is a process in which the same population of electrons is responsible for the synchrotron emission and self-absorption. In this model, the turnover occurs at a frequency at which the source becomes optically thick. Therefore, at higher frequencies, photons are seen from deep within the source and the intrinsic flux density is observed. However, at low frequencies, only emission coming from a thin shell at the surface of the source is visible, and emission from deeper within the source is absorbed, decreasing the total observed flux density. If we assume the region emitting (and absorbing) the synchrotron photons is homogeneous, we can model the spectrum as

\begin{equation}
S_{\nu} = a\left(\frac{\nu}{\nu_m}\right)^{-(\beta-1)/2}\left(\frac{1-e^{-\tau}}{\tau}\right),
\label{SSA_eqn}
\end{equation}

\noindent where $a$ is the normalisation parameter of the intrinsic synchrotron spectrum, $\nu_m$ is the turnover frequency, $\beta$ is the power-law index of the electron energy distribution, and $\tau$ is the optical depth given by $(\nu/\nu_m)^{-(\beta+4)/2}$. In this model, $\nu_m$ is the frequency at which the source becomes optically-thick, defined as the point at which the mean free path of electron-photon scattering is approximately the size of the source. This model predicts an optically-thick spectral index of 2.5 \citep{1981ARA&A..19..373K} and spectral indices shallower than this are generally attributed to inhomogeneity of the SSA regions, represented by multiple homogeneous SSA components.

\subsubsection{Free Free Absorption}

\label{sec_FFA}

The other dominant model used to account for the spectra of GPS and CSS sources is Free Free Absorption (FFA), which results from emission being attenuated by an ionized screen external to the emitting electrons. If a homogeneous screen surrounds the entire region of synchrotron emission, the spectrum is modelled by

\begin{equation}
S_{\nu} = a\nu^{\alpha}e^{-\tau_{\nu}},
\end{equation}

\noindent where $\alpha$ is the synchrotron spectral index, $\tau_{\nu}$ is the free-free optical depth, parametrized by $\tau_{\nu} = (\nu/\nu_0)^{-2.1}$, where $\nu_0$ is the frequency at which $\tau_{\nu} = 1$.

%

Another model, proposed by \citet{bic97}, assumes the screen is inhomogeneous, which is modelled by clouds with a power-law distribution of optical depths parametrized by $p$, such that the spectrum is given by

\begin{equation}
S_{\nu} = a(p+1)\gamma\left[p+1,\tau_{\nu}\right]\left(\frac{\nu}{\nu_0}\right)^{2.1(p+1)+\alpha},
\label{FFA}
\end{equation}

\noindent where $\gamma$ is the lower incomplete gamma function of order $p+1$, given by

\begin{equation}
\int_0^{\tau_{\nu}} e^{-x} x^p dx,
\end{equation}

\noindent and $\tau_{\nu} = (\nu/\nu_0)^{-2.1}$. 

As noted by \cite{2009AN....330..120F}, SSA will always occur to some degree in GPS and CSS sources where synchrotron emission is present. \citet{Orienti} suggests that SSA is responsible for the turnover in GPS and CSS sources, but that an additional contribution from FFA is detected in the most compact sources, such as in cases where the optically-thick spectral index is steeper than the SSA limit of 2.5 \citep{2008A&A...487..885O}. It is likely that both SSA and FFA are significant effects in the GPS and CSS population as a whole. However, since previous studies have generally lacked broad coverage of the spectra below the turnover, where the distinction between models is most significant \citep{Cal15}, the significance of the contribution of FFA is generally unknown. 

It is now much easier to comprehensively study the optically-thick spectra in GPS and CSS sources with low-frequency telescopes such as the Murchison Widefield Array \citep[MWA;][]{tin13} and the Low-Frequency Array \citep[LOFAR;][]{LOFAR}, and study the optically-thin spectra and break features with high-frequency radio telescopes with relatively large bandwidths, such as Australia Telescope Compact Array (ATCA). Using such telescopes to study the spectra of GPS and CSS sources has revealed that the \citet{bic97} inhomogeneous FFA model is consistent with the radio spectrum and other physical properties of several compact GPS sources \citep{Tin15,Cal15}. 

\subsubsection{Spectral breaks}

\label{break_sec}

Another feature of the spectra of GPS and CSS sources is the steepening of the spectral index at high frequencies, referred to as a spectral break. This effect, also known as spectral ageing, is due to synchrotron and inverse-Compton cooling in the jets and lobes, in which higher-energy electrons deplete more quickly, since their energy is expended faster. 

\cite{1962SvA.....6..317K} models the spectral break in a system in which the jets are continually switched on, injecting electrons into a volume with a constant magnetic field. In this model, there is an abrupt change in the spectrum at the break frequency, at which point the synchrotron spectrum steepens from $\alpha$ to $\alpha-0.5$, where $\alpha$ is the injection spectral index -- i.e. the synchrotron spectral index of fresh electrons, which is typically $-0.8$, given by $-(\beta - 1)/2$, where $\beta$ is the electron energy distribution. We refer to this break as \textit{continuous injection (CI)} break, and parameterise it as a multiplicative term given by

\begin{equation}
\left(\frac{\nu}{\nu_{\rm br}}\right)^{\displaystyle {\alpha - 0.5 + \frac{0.5}{1 + (\nu / \nu_{\rm br})^c}}}
\end{equation}

\noindent where $\nu_{\rm br}$ is the break frequency, and $c$ is a constant value determining the sharpness of the break, which we set to 5. For an optically-thin spectrum dominated by synchrotron emission, we expect the injection spectral index to be close to $\alpha = -0.8$. Therefore, if no CI break is observed in the data, and the optically-thin spectral index is $\alpha \gtrsim -0.8$, a CI break may exist above the range of the data. However, if no CI break is observed in the data, and the optically-thin spectral index is $\alpha \gtrsim -1.3$, a CI break may be below or hidden within the turnover \citep[e.g.][]{Cal15}. However, if the radio emission is dominated by the hotspot components, where particle acceleration is ongoing, a CI break may not be observable, or the spectral age may be underestimated \citep{1999A&A...345..769M}.

\cite{1973A&A....26..423J} describe an alternative model in which there is a momentary injection of relativistic electrons, causing a smooth exponential drop in flux density in the spectrum, given by $e^{-\nu/\nu_{\rm br}}$, where $\nu_{\rm br}$ is the exponential break frequency. We refer to this break as the \textit{exponential break}. 

Based on a source's break frequency, \citet{2003PASA...20...19M} derive its spectral age ($t_s$) based on the electron lifetime, by assuming negligible inverse Compton losses and an isotropization of the pitch angle following \cite{1973A&A....26..423J}, given by

\begin{equation}
t_s = 5.03 \times 10^4 \cdot B^{-1.5}[(1+z)\nu_{\rm br}]^{-0.5}~\mbox{years},
\label{spectral_age}
\end{equation}

\noindent where $B$ is the strength of the magnetic field in mG and $\nu_{\rm br}$ is the break frequency in GHz. 

A source whose jet has been continuously injecting electrons and then switches off, ceasing the injection of new electrons, produces a CI break followed by an exponential break \citep{1994A&A...285...27K}. For such a source, the exponential break frequency ($\nu_{br_{\rm exp}}$) relates to the CI break frequency ($\nu_{br}$) via

\begin{equation}
\nu_{br_{\rm exp}} = \nu_{br}\left(\frac{t_s}{t_{\rm off}}\right)^2,
\label{break_freqs}
\end{equation}

\noindent where $t_{\rm off}$ is the {\it turnoff time}, the time since the jet ceased injecting new electrons \citep{2007A&A...470..875P}.

\subsection{Turnover-linear size relation}

\label{turnover_linear_size}

The turnover frequency of GPS and CSS sources is observed to vary with the linear size as

\begin{equation}
\label{linear_size_turnover}
\log \nu_m = -0.21(\pm0.04) - 0.59(\pm0.05) \log l,
\end{equation}

\noindent where $l$ is the projected linear size of the radio source and $\nu_m$ is the intrinsic turnover frequency \citep{2014MNRAS.438..463O}. The small scatter around this linear fit shows that the there is a continuous rather than bimodal distribution, which implies that CSS sources are simply larger, older GPS sources. This is consistent with the great deal of overlap that exists between GPS and CSS sources, which are defined arbitrarily by their turnover frequencies. This relation indicates that the mechanism causing the peaked spectra is related to the source dimension. 

In the homogeneous SSA model, as the source expands, adiabatic expansion occurs in the mini-lobes that dominate the radio emission, causing their opacity to decrease, producing less SSA and therefore a lower-frequency turnover. Therefore, in the SSA model, this relation is well justified and indicates that the turnover frequency and linear size are both related to the age. Homogeneous FFA cannot account for this relation \citep{ODea98}. However, in the \citet{bic97} FFA model, the electron density within the external inhomogeneous medium decreases with distance from the core, allowing for the relation.

\subsection{Low-luminosity GPS and CSS sources}

Until recently, our understanding of GPS and CSS sources was limited to very bright Jy-level samples. Even now, their properties at faint levels are generally unknown. Amongst the faintest samples are the AT20G HFP samples \citep{2009AN....330..180H,2009MNRAS.397.2030H,2010MNRAS.408.1187H} and the Australia Telescope Large Area Survey \citep[ATLAS;][]{ATLAS_CDFS} CSS sample \citep{2012MNRAS.421.1644R}, which consist of mJy-level HFPs and sub-mJy CSS sources, respectively. The AT20G samples reach tens to hundreds of mJy and contain high-frequency turnovers that are much more subject to contamination by flat-spectrum quasars. The ATLAS CSS sample is much fainter, reaching sub-mJy levels, with a mean of $\sim$1 mJy, but only contains flux density measurements at two frequencies. 

\citet{2015arXiv151201851S} concluded that there is a large population of less-luminous GPS and CSS sources, which have so far eluded detailed study, due to the lack of sensitive large-area surveys at multiple frequencies, and the large time-requirement for characterising their morphologies with Very Long Baseline Interferometry (VLBI). \cite{2015MNRAS.448..252T} observed two such low-luminosity GPS and CSS sources with VLBI and proposed a luminosity-morphology break for compact radio galaxies analogous to the FR I/II break. \cite{2015arXiv151009061K} also concluded that GPS and CSS sources start to resemble mini FR I galaxies at low luminosity, which are the missing precursors of their larger-scale counterparts. However, very few low luminosity ($L_{\rm 1.4 GHz} < 10^{27}$ W Hz$^{-1}$) samples exist, especially those with broad spectral coverage and imaged with VLBI.

Examples of low-luminosity samples include \cite{2010MNRAS.408.2261K}, \cite{2014MNRAS.438..463O}, \citet{2015arXiv151201851S}, and \cite{2015arXiv151009061K}, which have typical luminosities of $10^{25-26}, 10^{25-30}, 10^{22-26}$, and $10^{23-27}$ W Hz$^{-1}$, respectively at 1.4, 0.375, 20 and 1.4 GHz. \cite{2014MNRAS.438..463O} bring together eight samples from the literature that have estimated turnover frequencies and linear sizes, which we use as a comparison. Despite these deep samples, \cite{2015arXiv151009061K} suggest that significant samples of low luminosity GPS and CSS sources are yet to be explored in deep radio surveys. Collier et al. (in prep.) present one such sample of 71 GPS and CSS sources with $L_{\rm 1.4 GHz} = 10^{21-27}$ W Hz$^{-1}$ using the deep radio observations of ATLAS, which consists of two of the deepest and most well-studied fields in the sky, as part of the broader project outlined in \cite{2015arXiv151107929C}. Here we present a subset of eight of the bright and compact sources from this sample of 71 that we have imaged with VLBI.

In this paper, we present a study of the radio spectra and high-resolution morphologies of a low-luminosity sample of GPS and CSS sources, using the radio data outlined in Section~\ref{ATLAS}. Our sample has 1.4 GHz flux densities between 3--119 mJy, and luminosities $L_{\rm 1.4 GHz} = 10^{23.5-26.5}$ W Hz$^{-1}$. Furthermore, each source has flux density measurements at 6--47 frequencies. Therefore, our sample represents one of the faintest and lowest luminosity samples that contains broad spectral coverage and mas VLBI imaging. 

We present the results in Section~\ref{VLBI_results}, in which we determine the properties of low-luminosity GPS and CSS sources, including their linear sizes and plausible absorption mechanisms. We use the radio spectra to test whether the spectra of GPS and CSS sources can be represented by FFA or the widely favoured homogenous SSA model with or without spectral breaks. We use all results to explore whether the properties of faint GPS and CSS sources are consistent with the well-known brighter samples. A discussion of individual sources is presented in Section~\ref{notes}, followed by a summary and conclusion in Section~\ref{VLBI_conclusions}. Throughout this paper, we use $\Omega_{\rm M} = 0.286$, $\Omega_{\rm \Lambda} = 0.714$, and $H_0 = 69.6$ km s$^{-1}$ Mpc$^{-1}$.

\section{Radio Data}

\label{ATLAS}

To select the Collier et al. (in prep) ATCA sample, and our VLBI sample, we started with the 1.4, 1.71 and 2.3~GHz radio observations from ATLAS \citep{ATLAS_Zinn,ATLAS_DR3}, which covers $\sim$7 square degrees in the Chandra Deep Field South \citep[CDFS;][]{2002ApJ...566..667R} and the European Large Area ISO Survey South 1 \citep[ELAIS-S1;][]{2000MNRAS.316..749O} down to an r.m.s. of $\sim$15~$\mu$Jy beam$^{-1}$ at 1.4~GHz. The third data release \citep[DR3;][]{ATLAS_DR3} contains $5\: 118$ sources, and spectroscopic redshifts for $\sim$30 per cent of the sources from the OzDES Global Reference Catalogue \citep[][Childress et al. submitted to MNRAS]{2015MNRAS.452.3047Y}.

The DR3 catalogue consists of a 1.4 GHz flux density and a spectral index, originally derived between two sub-bands at 1.4 and 1.71 GHz. We recover the 1.71 GHz flux density and uncertainty as:

\begin{equation}
S_{\rm 1.71 GHz} = S_{\nu_{\rm obs}}x^{\alpha}, {\rm and}
\end{equation}

\begin{equation}
\delta S_{\rm 1.71 GHz} = \sqrt{(x^{\alpha})^2 + (S_{\nu_{\rm obs}}\ln(x)x^{\alpha}\delta\alpha)^2},
\end{equation}

\noindent where $x = {\rm 1.71} / {\nu_{\rm obs}~\rm (GHz)}$. Due to its relatively small bandwidth, for some sources, the 1.71 GHz sub-band flux density has a large uncertainty, while the full-band 1.4 GHz flux density has a small uncertainty due to its relatively large bandwidth.

While selecting our samples, we also used deep 150 and 325 MHz Giant Metre-wave Radio Telescope (GMRT) observations \citep{2009MNRAS.395..269S}, 843 MHz Molonglo Observatory Synthesis Telescope (MOST) observations of the ELAIS-S1 field \citep{2012MNRAS.421.1644R}, and 5.5~GHz ATCA observations of the ECDFS \citep{2012MNRAS.426.2342H,2015MNRAS.454..952H}, which covers 0.25 deg$^2$ of the CDFS.

From these data, Collier et al. (in prep) selected the faintest GPS/CSS sample to date and undertook high-resolution observations using the new 4~cm receiver on the ATCA (project ID: C2730), observing 71 sources at 5.5 and 9.0 GHz down to r.m.s.~levels between tens and hundreds of $\mu$Jy beam$^{-1}$, depending on the strength of each source. More details on the observations, reduction, and analysis will be presented in Collier et al. (in prep). 

\subsection{Selection Criteria}

As we were not trying to select a complete sample, we did not apply rigorous selection criteria, but selected a number of interesting GPS and CSS candidates from the ATLAS fields based on the data available at the time, so our selection criteria characterise the sources as a whole:

\begin{enumerate}
\item Has a peaked radio spectrum with $\nu_m \sim 1$ GHz (GPS); or
\item Has a steep radio spectrum with $\alpha \lesssim -0.7$ (CSS); and
\item Is compact (handled separately for different fields, since the Collier et al. (in prep) data of the CDFS were not available at the time of selection):
\begin{enumerate}
\item[a)] ELAIS-S1 source unresolved in Collier et al. (in prep.) 9~GHz data ($\theta \lesssim 1\arcsec$; $l \lesssim$ 6 kpc at $z=0.5$);
\item[b)] ECDFS source unresolved in \citet{2012MNRAS.426.2342H} 5.5~GHz data ($\theta \lesssim 2\arcsec$; $l \lesssim$ 12 kpc at $z=0.5$);
\item[c)] CDFS source with mas-scale size derived from Equation~\ref{linear_size_turnover}.
\end{enumerate}
\end{enumerate}

\noindent The first two criteria were based on visual inspection of plots of the radio spectrum, which used all available flux densities, measured from beam-matched images where possible. The third criterion ensured that the sources were sufficiently compact for VLBI observations, so they would yield a high enough signal-to-noise (S/N). We selected the Collier et al. (in prep) sample in this way, and from this sample, selected eight of the best sources for VLBI that could be observed within the time that we were allocated, selected in the same way and also using the data from Collier et al. (in prep), with five in the ELAIS-S1 and three in the CDFS. Collier et al. (in prep) later observed all CDFS sources with ATCA at 5.5 and 9.0 GHz, except for the one source from the ECDFS (CI0020), which had already been observed at these frequencies by \citet{2012MNRAS.426.2342H}.

\subsection{VLBI observations}

\label{obs}

We observed our sample with the Australian Long Baseline Array (LBA; project ID: V506) over two days, starting with the ELAIS-S1 sources on 2013 November 21 (V506a), and the CDFS sources on 2014 February 21 (V506b). For V506a, the array consisted of the CSIRO telescopes of the Australian Square Kilometre Array Pathfinder \citep[ASKAP;][]{2007PASA...24..174J,2009IEEEP..97.1507D}, ATCA, and Parkes, in addition to the University of Tasmania telescopes of Hobart and Ceduna (At-Ak-Pa-Ho-Cd). For V506b, the CSIRO telescope Mopra (Mp) was added to the array. This gave resolutions as high as $\sim$15 mas. In both cases, observations were made at a central frequency of 1.634 GHz with a 64 MHz bandwidth at each of the two circular polarizations (left and right; L and R) and were obtained over 10 hour periods. The observations were structured to cycle between the targets and nearby calibrators, with scan lengths of 90 seconds.

\subsection{Data reduction}

\label{VLBI_reduction}

\subsubsection{Calibration}

\begin{figure}
\begin{center}
\includegraphics[angle=-90,width=0.4\textwidth]{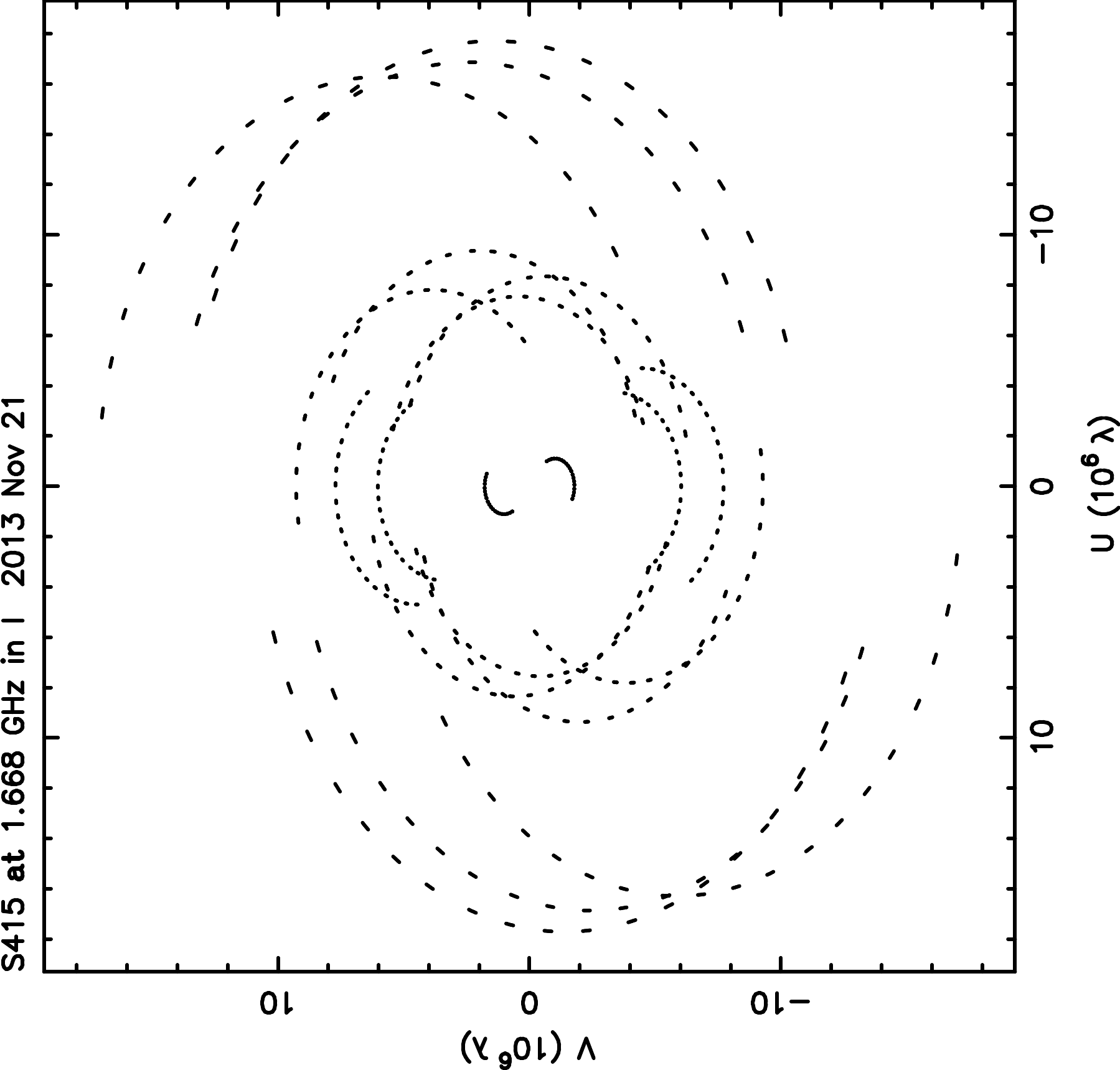}
\includegraphics[angle=-90,width=0.4\textwidth]{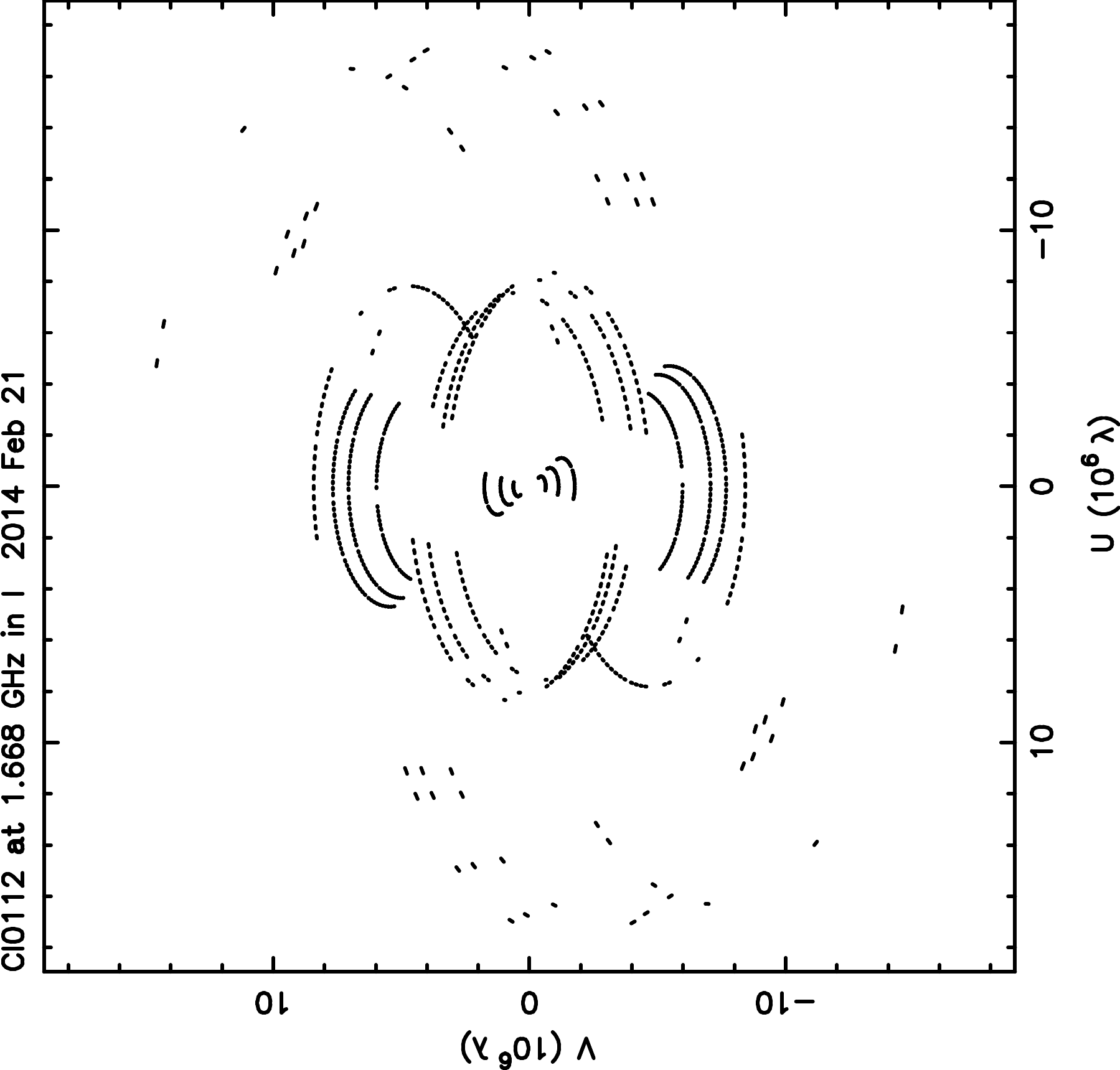}
\caption{The typical $(u,v)$ coverage achieved during the first (top) and second (bottom) days of our LBA observations.}
\label{LBA_uv_cov}
\end{center}
\vspace{-5mm}
\end{figure}

The LBA data for all eight sources were correlated at Curtin University using the DiFX software correlator \citep{2007PASP..119..318D,2011PASP..123..275D} with 128 channels across the 64 MHz band and 2 s integration times. Only parallel hand polarizations were correlated (RR and LL). The typical $(u,v)$ coverage achieved on the first and second days are shown in Fig.~\ref{LBA_uv_cov}. The visibility data were imported into \textsc{aips} for standard processing, as briefly described below.

The visibility data were calibrated in amplitude using measured system temperatures for each telescope from the time of observation, as well as known gains for each telescope. The resulting amplitudes for the calibrators (which were unresolved) were then compared to flux densities measured from the ATCA. Adjustments to the telescope gains were made, with a subsequent refinement of the calibrated amplitudes. We estimate the flux density scale is accurate to within $\sim$10 per cent uncertainty.

The visibilities for the calibrator sources were then fringe-fitted to solve for delays and phases for each antenna, interpolated to the times of the target observations, and applied to the target visibilities. This standard phase referencing step calibrates the visibility phases for the target sources, allowing imaging to proceed.

\subsubsection{Imaging and source extraction}

The target visibility data were exported from \textsc{aips} into \textsc{difmap} for further analysis. Two sources yielded no detections in the $(u,v)$ plane, and were not imaged. For the remaining six, because of the relatively sparse nature of the measurements (see Fig.~\ref{LBA_uv_cov}), a model-fitting approach was adopted, using the task \texttt{MODELFIT} in \textsc{difmap}. For each target source, a model for the source structure was generated using the smallest number of Gaussian components required to fit the visibilities. The models were iteratively fit to the data and used for self-calibration (when the sources were detected with high enough S/N). Four sources (CI0020, s895, s415, and s150) were imaged with a pixel size of $\sim$3 mas, while the other two (CI0008 and CI0112) were imaged with a pixel size of $\sim$6 mas.

\subsection{Ancillary radio data}

In addition to the data from which our sample was selected, we made use of newer GMRT observations of the ELAIS-S1 field at 150 \citep{Intema16} and 610 MHz (Intema et al., in prep), and of the CDFS field at 150 MHz (PI: Shankar). We also made use of the large ensemble of deep radio data available in the CDFS field, including MOST observations at 408 and 843 MHz (PI: Hunstead; Crawford et al., in prep.), ATCA observations at 5.5, 9, 18 and 20 GHz from the AT20G pilot survey and follow-up program \citep{2014MNRAS.439.1212F}, and ATCA 34 GHz observations of a strip within the ECDFS (project ID C2317; PI: Beelen).

\subsubsection{ASKAP}

We also used ASKAP observations of $\sim$19 deg$^2$ of the CDFS that was made at 844 MHz using the Boolardy Engineering Test Array \citep[BETA;][]{2014PASA...31...41H,2016PASA...33...42M}. BETA was a prototype of ASKAP that used six of the 36 dishes, each equipped with the first generation (Mark I) of phased array feeds. To image the CDFS using BETA, the ASKAP Commissioning and Early Science (ACES) team followed a similar reduction process to \cite{2016MNRAS.457.4160H}, and achieved an r.m.s. of $\sim$640~$\mu$Jy beam$^{-1}$ and a resolution of $91\arcsec \times 56\arcsec$.

\subsubsection{GLEAM}
\label{GLEAM}

Lastly, we used MWA data from the Galactic and Extragalactic All-sky MWA \citep[GLEAM;][]{GLEAM} extragalactic catalogue of the sky south of $\delta=30$ degrees \citep{GLEAM_DR1}. The GLEAM catalogue is based on sources detected at 200 MHz within a deep image covering $170-231$~MHz. Sources detected in this image were used as {\it a-priori} information by the \textsc{aegean} source finder \citep{Aegean}, via its priorized fitting feature\footnote{\href{https://github.com/PaulHancock/Aegean/wiki/Priorized-Fitting}{https://github.com/PaulHancock/Aegean/wiki/Priorized-Fitting}}, to extract 20 sub-band flux densities, each of which had 7.68~MHz bandwidths. Therefore, all sources contain measurements at all 20 sub-bands, although many are very low in S/N.
 
Since most of the sources we selected were faint, we averaged together three sets of four of the 7.68~MHz GLEAM sub-bands between $72-103$, $103-134$, and $139-170$~MHz. We derived the average flux densities as

\begin{equation}
\frac{\sum S_{\nu}}{N} \pm \frac{1}{\sqrt{N}}\sqrt{\sum(\delta S_{\nu}^2)},
\end{equation}

\noindent where $\delta S_{\nu}$ is the uncertainty on the flux density $S_{\nu}$ at frequency $\nu$, and N is the number of 7.68~MHz channels that we averaged together, which was four. In combination with the deep $170-231$~MHz measurement, this resulted in four GLEAM flux density measurements, which we used for three faint sources detected in the deep image (CI0020, s895, and s415). If the uncertainty was larger than the flux density, we discarded the measurement.

\subsubsection{Source extraction}

We used \textsc{pybdsm} \citep{pybdsm} to perform source extraction for some of the ancillary radio data. \textsc{pybdsm} calculates the background r.m.s. and mean maps, identifies islands of emission, fits multiple Gaussians to each island, derives the residuals, and groups Gaussians into sources. It then performs further source extraction on lower resolution images, generated by processing the residual images with an {\it \'a trous} wavelet transformation, at the end of which, a Gaussian catalogue and a source catalogue are written. Since we were interested in the total sum of flux densities over all components, we grouped all Gaussians belonging to an island into one source, and used an island threshold of 2.0, allowing \textsc{pybdsm} to flood-fill adjacent pixels down to 2$\sigma$. Additionally, we used an adaptive r.m.s. box, which allowed for higher r.m.s. values due to artefacts close to strong sources.

Source extraction was performed on the images from GMRT at 150, 325, and 610 MHz, MOST at 408 and 843 MHz, ASKAP at 844 MHz, and ATCA at 34 GHz. We also performed source extraction on the ATLAS 2.3 GHz images for a few sources that had been omitted from the \cite{ATLAS_Zinn} catalogue. In the ELAIS-S1 field, \textsc{pybdsm} had already been used to perform source extraction on the 610 MHz mosaic (Intema et al., in prep.).

\vspace{-5mm}

\section{Results}

\label{VLBI_results}

\begin{figure*}
\begin{center}
\vspace{-2mm}
\hspace{-12mm}
\includegraphics[trim=0mm 7mm 30mm 0mm,width=0.33\textwidth]{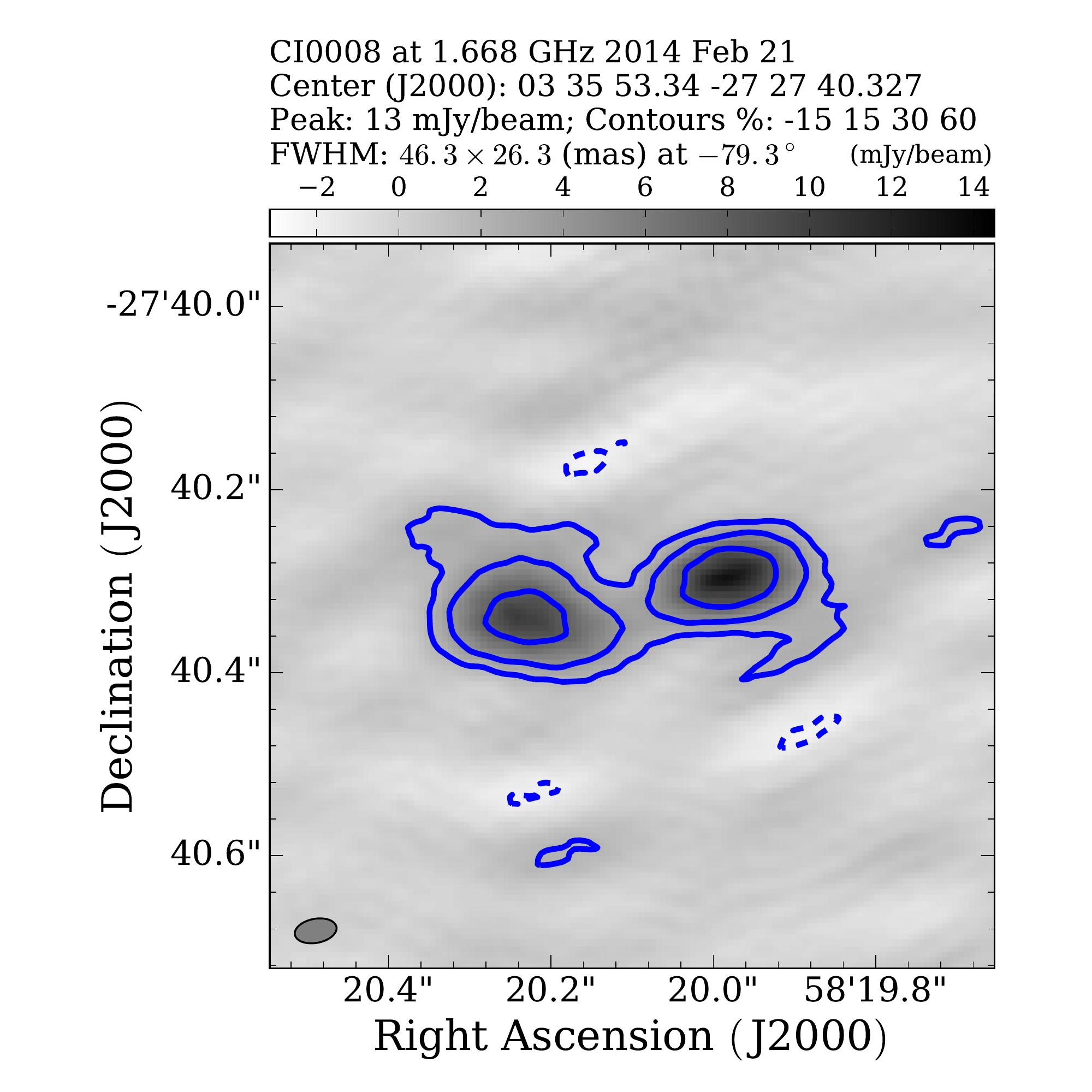}\includegraphics[trim=0mm 7mm 30mm 0mm,width=0.33\textwidth]{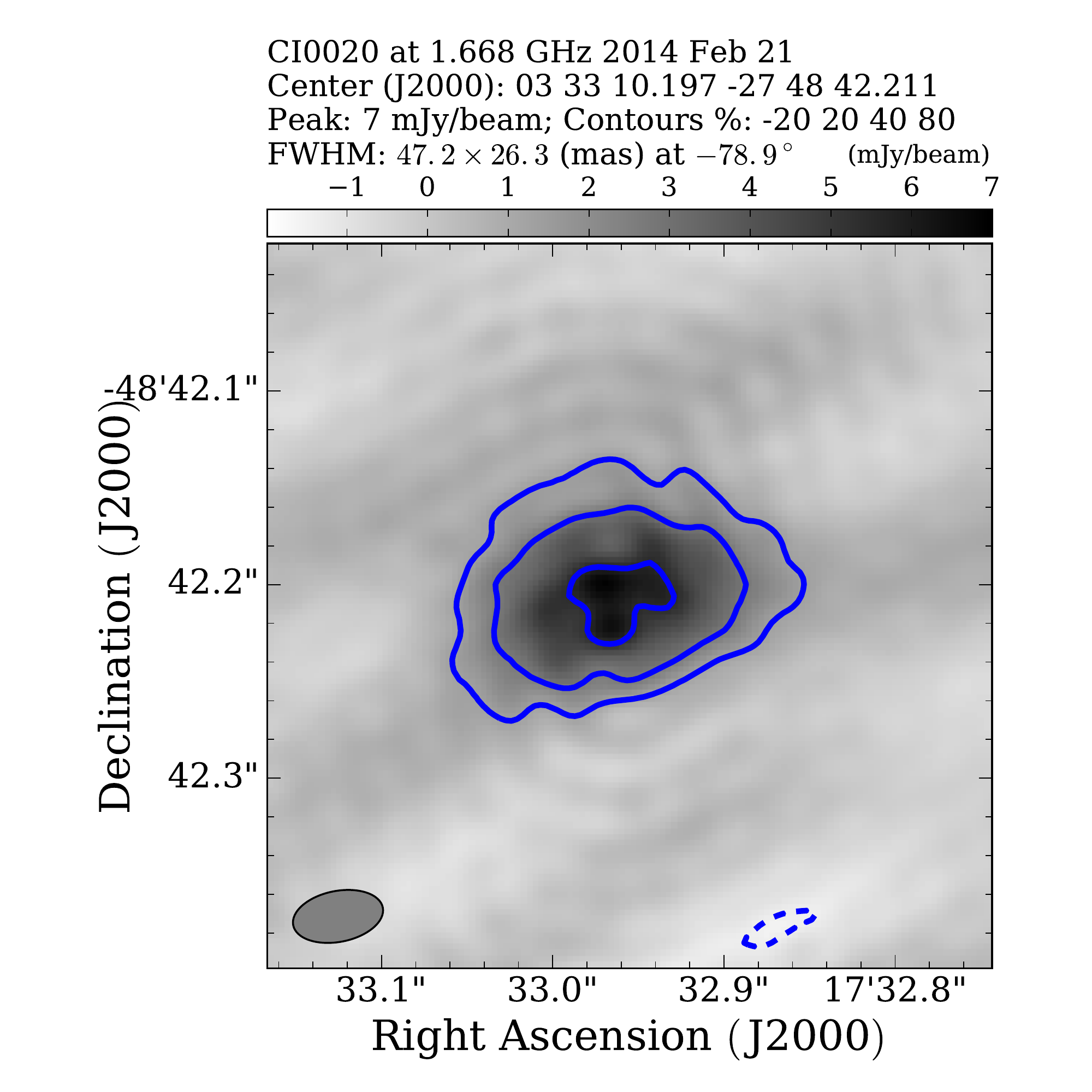}\includegraphics[trim=0mm 7mm 30mm 0mm,width=0.33\textwidth]{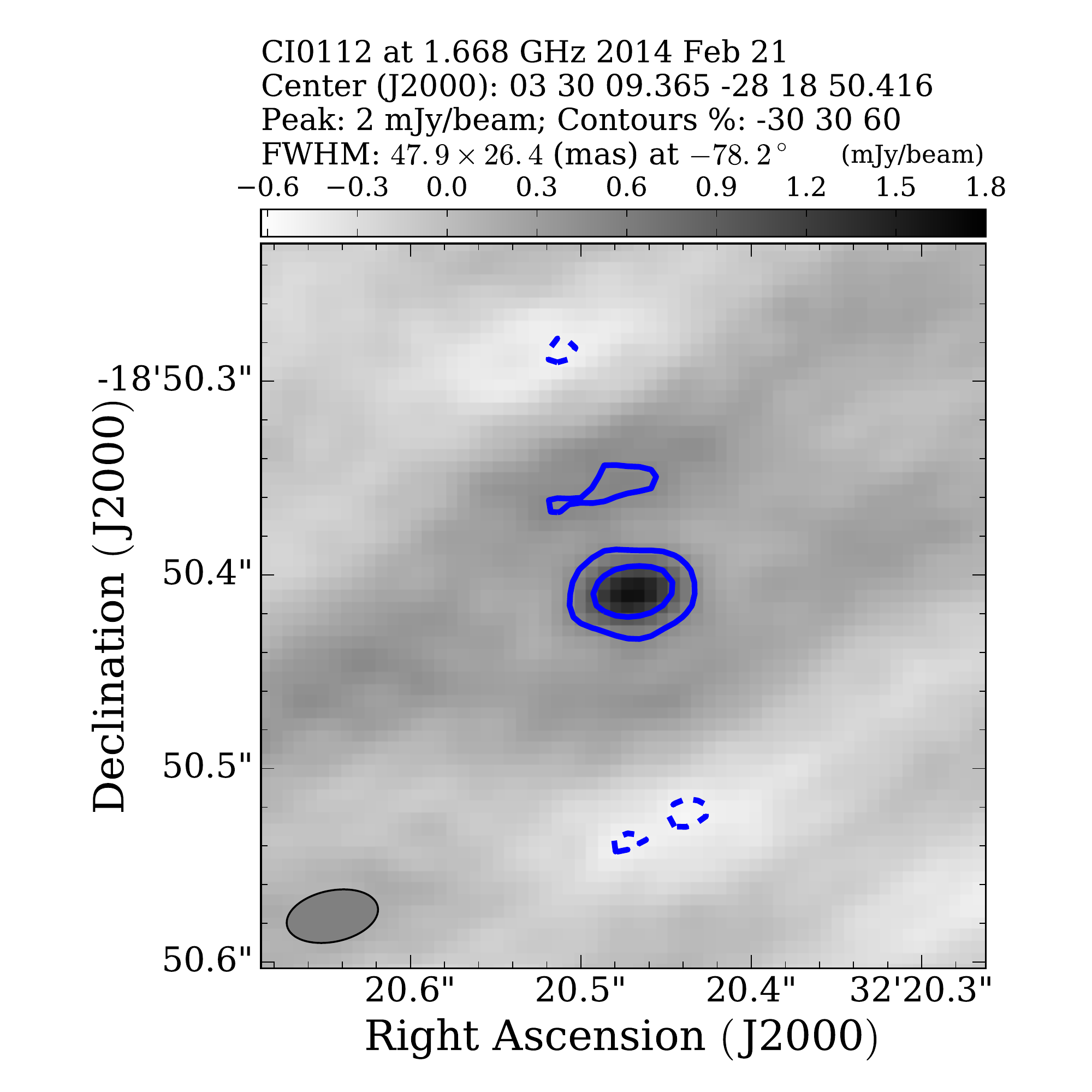}\\
\hspace{-12mm}
\includegraphics[trim=0mm 7mm 30mm 0mm,width=0.33\textwidth]{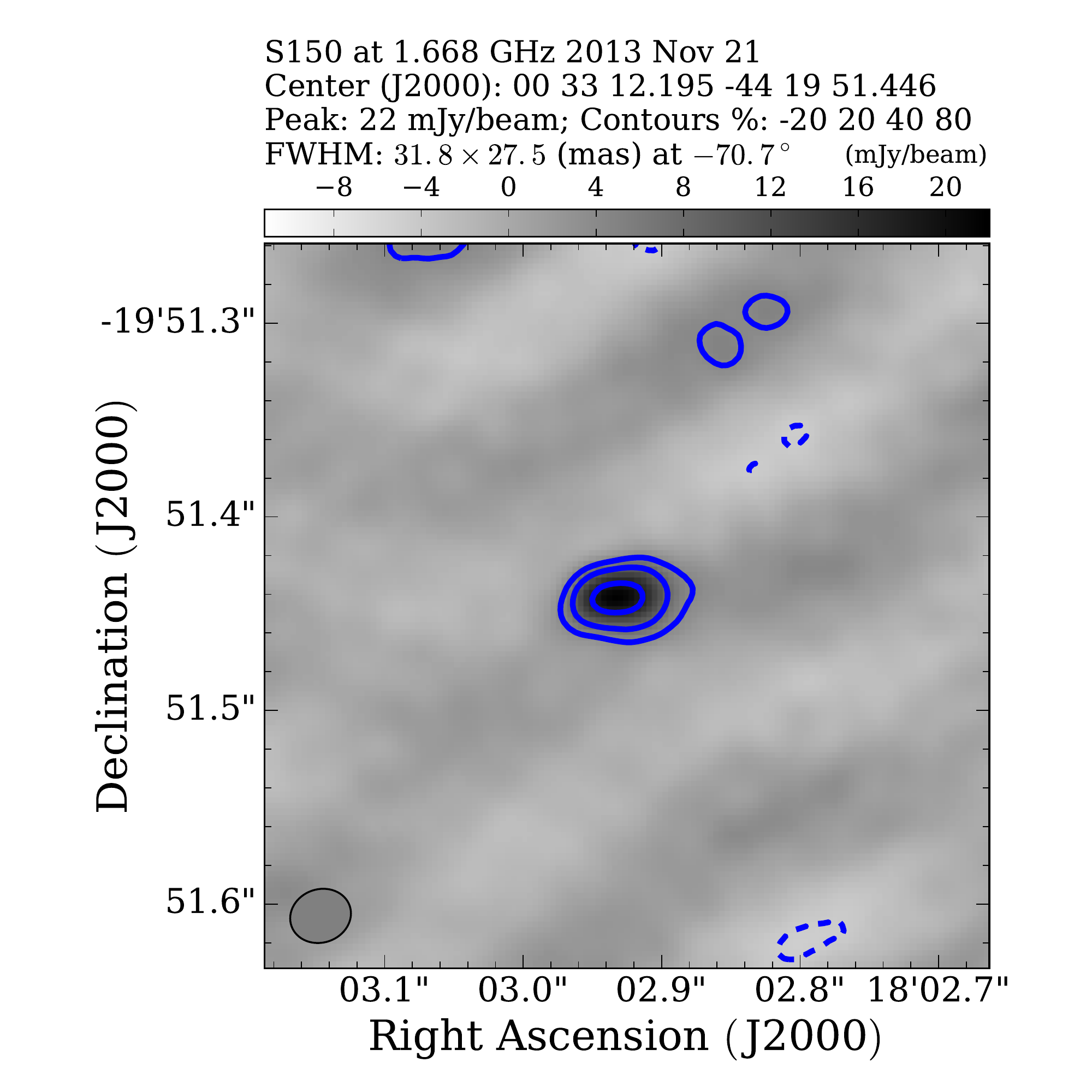}\includegraphics[trim=0mm 7mm 30mm 0mm,width=0.33\textwidth]{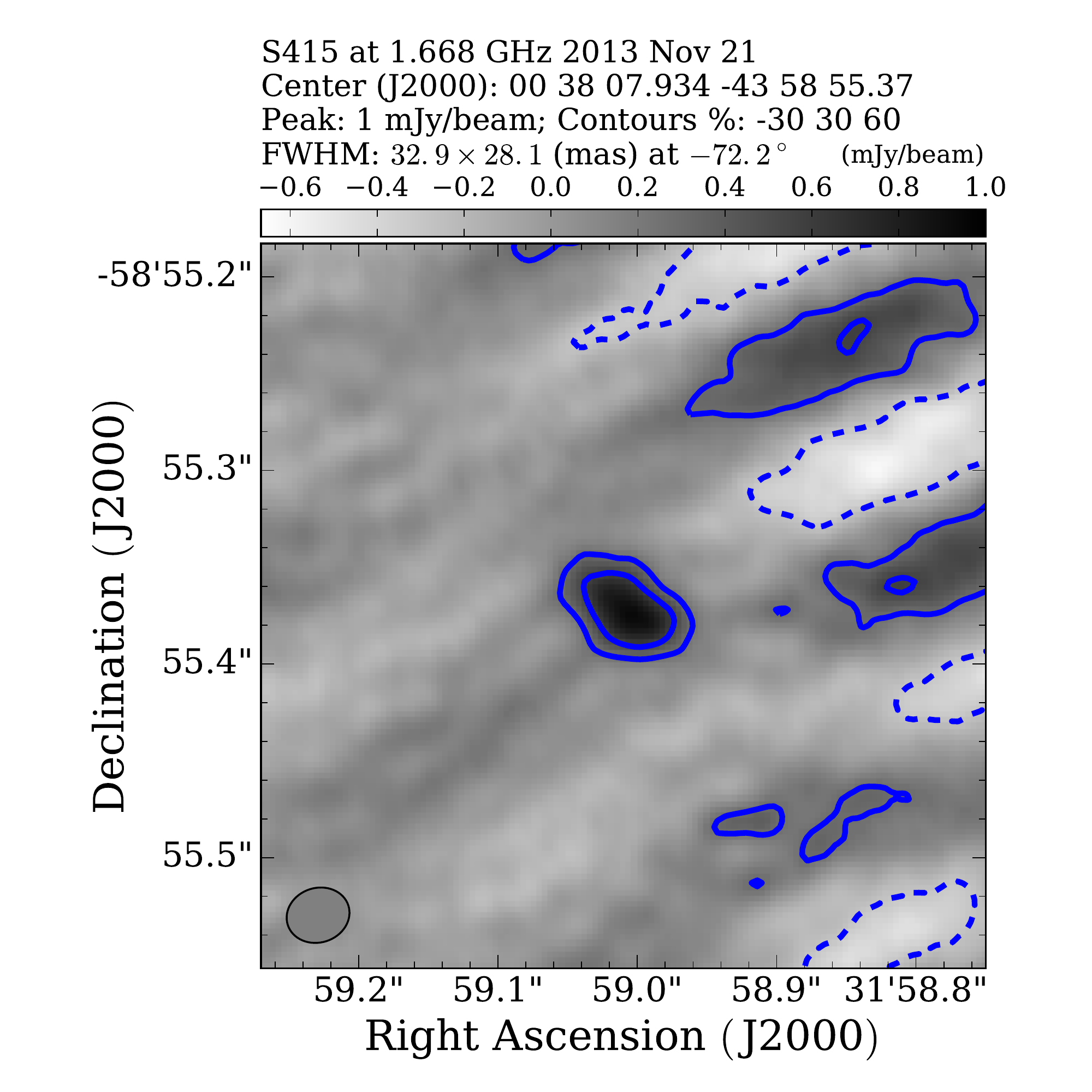}\includegraphics[trim=0mm 7mm 30mm 0mm,width=0.33\textwidth]{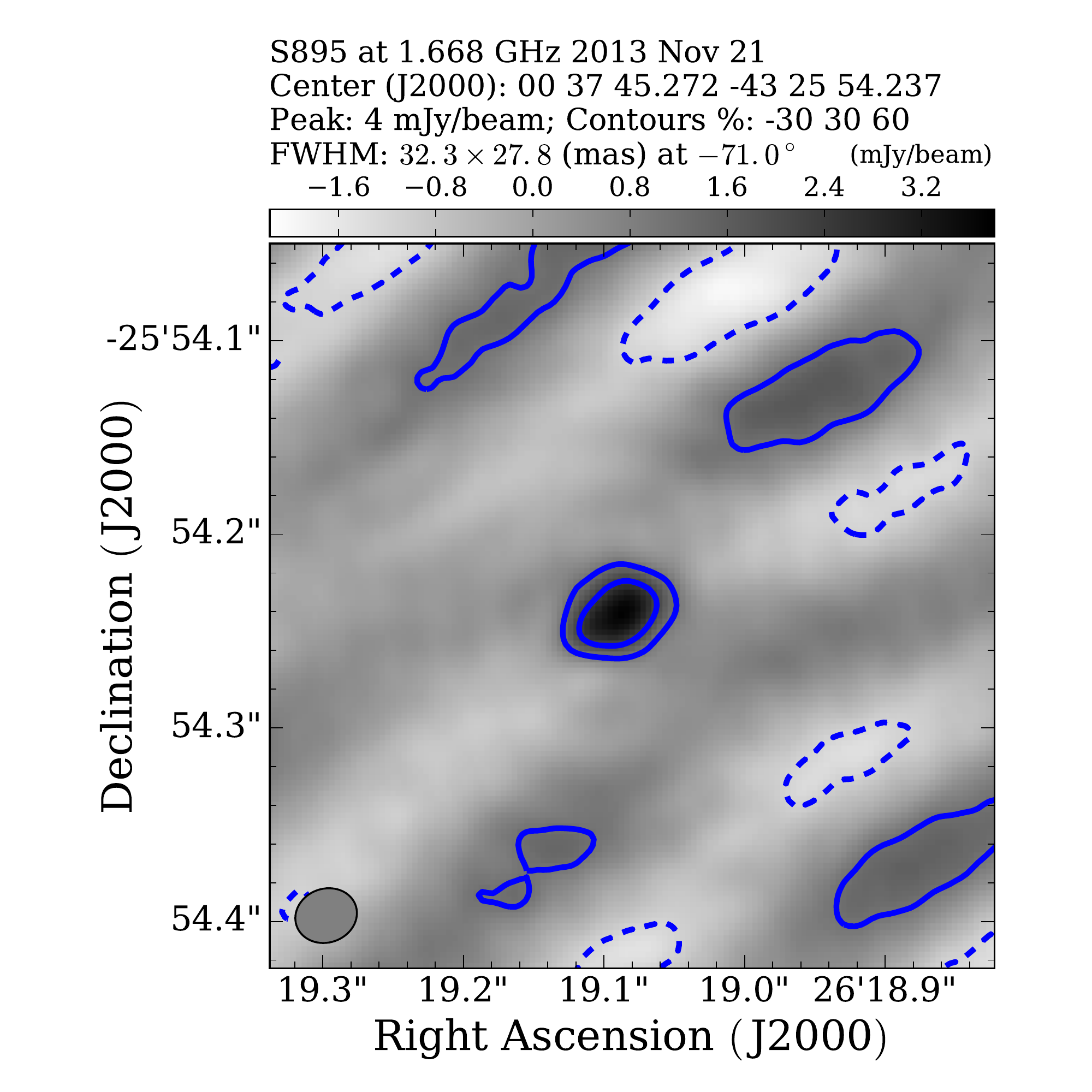}
\caption{The LBA images of the six detected GPS and CSS candidates. The contours represent the displayed percentages of the image peak, including a negative dashed contour. The synthesised beam is shown by the shaded ellipse in the bottom left hand corner, and its size is displayed above the map.}
\label{VLBI_imgs}
\end{center}
\end{figure*}

The final models fit to the sources detected with the LBA are listed in Table~\ref{VLBI_models}, and the upper limits for the undetected sources are listed in Table~\ref{VLBI_non_detections}. The corresponding images are shown in Fig.~\ref{VLBI_imgs}.

Large LBA-ATCA flux density ratios exist amongst the two most resolved sources, CI0008 and CI0020, as shown in Table~\ref{VLBI_models}. Given the sparse $(u,v)$ coverage (see Fig.~\ref{LBA_uv_cov}), particularly on the short baselines, these large ratios are most likely due to poorly constrained model component amplitudes. However, the VLBI observations are not used in the spectral models, but are used primarily to measure their linear sizes, which are not affected by this issue. The particular lack of short baselines on the first day may account for the two non-detections, which may have been resolved out.

\subsection{Linear sizes}

For single component sources, the largest linear size (LLS) was derived from the major axis of the fitted model. Source CI0008 was resolved into two components, so we derived the LLS from the maximum angular separation between the components, given by

\begin{equation}
\Theta + \theta_{\rm gauss1} + \theta_{\rm gauss2} - \theta_{\rm psf},
\label{ang_sep}
\end{equation}

\noindent where $\Theta$ was the separation between the Gaussians derived from their RA and Dec, and $\theta_{\rm gauss1}$, $\theta_{\rm gauss2}$ and $\theta_{\rm psf}$ were respectively the radii of the first and second Gaussians and the synthesised beam at the position angle (PA) subtended between the two Gaussians ($\rm PA_{sky}$), defined as

\begin{equation}
\theta = \sqrt{(a \cos({\rm PA_{sky} - PA_{gauss}}))^2 + (b \sin({\rm PA_{sky} - PA_{gauss}}))^2},
\end{equation}

\noindent where $a$, $b$ and $\rm PA_{gauss}$ were respectively the FWHM of the major and minor axes, and the PA of the Gaussian, either the \textsc{pybdsm} Gaussian, or the synthesised beam. $\theta_{\rm psf}$ was simply taken as the major axis of the FWHM, since ${\rm PA_{gauss} - PA_{sky}} < 1$~degree.

\vspace{-3mm}

\subsection{Variability}

For sources CI0008, CI0020, s895 and s415, we used overlapping or very close frequency measurements from different epochs separated by a number of years, enabling us to constrain their variability. These included GMRT/MWA measurements at $\sim$150 MHz and MOST/ASKAP measurements at $\sim$843 MHz, all of which agreed within the uncertainties. 

For source CI00020, we also used the \cite{2014MNRAS.439.1212F} follow up observations of the AT20G pilot survey at 5.5 and 9 GHz,

\begin{landscape}
\begin{center}

\begin{table}
\centering
\caption{The GPS and CSS candidates detected with the LBA. Shown is the ID, the fitted RA, Dec, 1.67 GHz flux density, r.m.s., S/N, major axis, minor axis and PA from the LBA image, the 1.71 GHz ATCA flux density derived (see Section~\ref{ATLAS}) from ATLAS DR3 \protect\citep{ATLAS_DR3}, the LBA-ATCA flux density ratio, the redshift and its reference, the largest linear size (LLS), and the 1.4 GHz luminosity \protect\citep{ATLAS_DR3}. All values are given to two significant figures. Source CI0008 was fit with two components, which are listed separately. The LBA-ATCA flux density ratio listed for this source represents the sum of the flux density of both components as a fraction of the 1.71 GHz ATCA flux density. The LLS of this source corresponds to the distance between these components, measured from Equation~\ref{ang_sep}. The LLS for all other sources is derived from the major axis of the FWHM. Sources with ID prefix `s' are from the ELAIS-S1 field, following the source IDs from \protect\cite{ATLAS_ELAIS}. Sources with ID prefix `C' are from the CDFS field, following an earlier version of the ATLAS DR3 catalogue. References for the redshifts are listed as the following: (1) = photo$-z$ from \protect\cite{Rowan-Robinson2008}; (2) = \protect\cite{2012MNRAS.426.3334M}; (3) = \protect\cite{2011ApJ...741....8C}; (4) = \protect\cite{2008ApJS..179...95M}.}
\label{VLBI_models}
\sisetup{round-mode=figures}
\begin{tabular}{ccc
S[round-precision=2]
S[round-precision=2]
S[round-precision=2]
c
S[round-precision=2]
c
S[round-precision=2]
S[round-precision=2]
c
S[round-precision=2]
S}
\hline
    \multicolumn{1}{c}{ID} & 
    \multicolumn{1}{c}{RA} & 
    \multicolumn{1}{c}{Dec} & 
    \multicolumn{1}{c}{$S_{\rm LBA}\dagger$} & 
    \multicolumn{1}{c}{r.m.s.} & 
    \multicolumn{1}{c}{S/N} & 
    \multicolumn{1}{c}{$\Theta_{\rm maj} \times \Theta_{\rm min}$} & 
    \multicolumn{1}{c}{PA} &
    \multicolumn{1}{c}{$S_{\rm ATCA}$} & 
    \multicolumn{1}{c}{$S_{\rm LBA}$}$\dagger$ & 
    \multicolumn{1}{c}{$z$} &
    \multicolumn{1}{c}{Ref.} &
    \multicolumn{1}{c}{LLS} &
    \multicolumn{1}{c}{$L_{\rm 1.4 GHz}$} \\

    \multicolumn{1}{c}{} & 
    \multicolumn{2}{c}{(J2000)} & 
    \multicolumn{1}{c}{(mJy)} & 
    \multicolumn{1}{c}{(mJy/beam)} & 
    \multicolumn{1}{c}{} & 
    \multicolumn{1}{c}{(mas $\times$ mas)} & 
    \multicolumn{1}{c}{(deg)} &
    \multicolumn{1}{c}{(mJy)} & 
    \multicolumn{1}{c}{$\overline{S_{\rm ATCA}}$} & 
    \multicolumn{1}{c}{} &
    \multicolumn{1}{c}{} &
    \multicolumn{1}{c}{(kpc)} &
    \multicolumn{1}{c}{W Hz$^{-1}$} \\
\hline
  s150 & 00:33:12.1954 & $-$44:19:51.4418 & 29.5627 & 0.194863 & 151.710175867147 & 29 $\times$ 0.0 & -86.0581 & 17 $\pm$ 1.0 & 1.70392987167517\\
  s895 & 00:37:45.2726 & $-$43:25:54.2412 & 6.49979 & 0.0879097 & 73.9371195670102 & 33 $\times$ 18 & -58.9354 & 6.4 $\pm$ 0.85 & 1.00922155578357 & 1.137962 & (1) & 0.27868983499739 & 5.9E25\\
  s415 & 00:38:07.9339 & $-$43:58:55.3721 & 1.95128 & 0.0859764 & 22.6955304013659 & 47 $\times$ 14 & 52.9957 & 6.0 $\pm$ 0.88 & 0.32548790694365 & 0.50660002231597 & (2) & 0.29033771507890 & 7.3E24\\
  CI0008-1 & 03:35:53.3319 & $-$27:27:40.2979 & 104.178999999999 & 0.179954 & 578.920168487502 & 130 $\times$ 74 & -83.6736 & 99 $\pm$ 1.8 & 2.3561788274 & 0.2558 & (2) & 1.79497200042710 & 2.4E25\\
  CI0008-2 & 03:35:53.3486 & $-$27:27:40.3363 & 128.758 & 0.179954 & 715.505073518788 & 150 $\times$ 99 & 70.8773 & 99 $\pm$ 1.8 & 2.3561788274 & 0.25580000877380 & (2) & 1.79497204578456 & 2.4E25\\
  CI0112 & 03:30:09.3647 & $-$28:18:50.4100 & 1.86864 & 0.0831565 & 22.4713642349064 & 17 $\times$ 0.0 & 38.481 & 2.3 $\pm$ 0.91 & 0.8187146991732 & 0.28701654076576 & (3) & 0.07303321263902 & 6.9E23\\
  CI0020 & 03:33:10.1976 & $-$27:48:42.2056 & 47.9792 & 0.10353 & 463.432821404423 & 110 $\times$ 75 & -75.2002 & 21 $\pm$ 1.0 & 2.33623532956857 & 1.02900004386901 & (4) & 0.91916463824882 & 1.3E26\\
\hline
\end{tabular}
\\$\dagger$ See Section~\ref{VLBI_results} for a discussion about the uncertainty on the VLBI flux densities.
\end{table}

\vspace{20mm}

\begin{table}
\centering
\caption{The GPS and CSS candidates not detected with the LBA. Shown is the ID, RA, Dec and 1.4 GHz ATCA flux density from ATLAS DR1 \protect\citep{ATLAS_ELAIS}, the r.m.s. and 1.67 GHz flux density upper limit from the LBA, the redshift and the 1.4 GHz luminosity. All values are given to two significant figures. Both sources are from the ELAIS-S1 field, and use the source IDs and 1.4 GHz flux densities from DR1, since one source is outside the field catalogued in DR3 by \protect\cite{ATLAS_DR3}. We quote a 6.75$\sigma$ flux density upper limit following the approach from \protect\cite{Deller2014}. Both redshifts are photometric redshifts from \protect\cite{Rowan-Robinson2008}.}
\label{VLBI_non_detections}
\sisetup{round-mode=figures}
\begin{tabular}{ccccc
S[round-precision=2]
S[round-precision=2]
c
S[round-precision=2]}
\hline
    \multicolumn{1}{c}{ID} & 
    \multicolumn{1}{c}{RA} & 
    \multicolumn{1}{c}{Dec} & 
    \multicolumn{1}{c}{$S_{\rm ATCA}$} & 
    \multicolumn{1}{c}{$S_{\rm LBA}$} & 
    \multicolumn{1}{c}{r.m.s.} & 
    \multicolumn{1}{c}{$z$} &
    \multicolumn{1}{c}{Ref.} &
    \multicolumn{1}{c}{$L_{\rm 1.4 GHz}$} \\

    \multicolumn{1}{c}{} & 
    \multicolumn{2}{c}{(J2000)} & 
    \multicolumn{1}{c}{(mJy)} & 
    \multicolumn{1}{c}{(mJy/beam)} &
    \multicolumn{1}{c}{(mJy/beam)} & 
    \multicolumn{1}{c}{} &
    \multicolumn{1}{c}{} &
    \multicolumn{1}{c}{W Hz$^{-1}$} \\
\hline
  s798 & 00:39:07.934 & $-$43:32:05.833 & 7.8 $\pm$~0.05 & $<$ 0.55 & 0.0806853 & 0.399587 & (1) & 4.47E24\\
  s1218 & 00:35:08.380 & $-$43:00:04.202 & 33 $\pm$~0.11 & $<$ 1.3 & 0.193977 & 0.629296 & (1) & 5.69E25\\
\hline
\end{tabular}
\end{table}

\end{center}
\end{landscape}

\noindent both of which agreed within the uncertainties with the nearby measurements. Using the 18 and 20 GHz observations separated by more than 3 years, \cite{2014MNRAS.439.1212F} estimated a 2.6 per cent variability index for source CI0020, concluding it was non-variable. 

Following \cite{2014MNRAS.439.1212F}, we test for variability using a $\chi^2$-test, where the value is given by 

\begin{equation}
\label{v}
\chi^2 = \frac{(S_1 - S_2)^2}{\sigma_1^2 + \sigma_2^2},
\end{equation}

\noindent where $S_1$, $S_2$, $\sigma_1,$ and $\sigma_2$ are the flux density and uncertainty of the nearby or overlapping frequencies $\nu_1$ and $\nu_2$. As in \cite{2014MNRAS.439.1212F}, we classify sources as non-variable when the null hypothesis that $S_1$ and $S_2$ are the same is supported with a probability of 1 per cent or greater, given by $\chi^2 < 6.63$. Table~\ref{var_tab} shows the $\chi^2$ values we calculate for all overlapping or nearby frequencies, from which we conclude that these four sources are not variable.

Since our radio spectra span such a large range of frequencies over a large range of epochs, we expect variable sources will not maintain a typical GPS or CSS spectrum over these epochs. This appears to be the case with source CI0112, which gave a flat spectrum based on the simultaneous 5.5 and 9.0 GHz measurements, and which may be affected by variability. However, for source s150, we cannot rule out a low level of variability that does not significantly affect the spectral shape.

\begin{table}
\begin{center}
\caption{The $\chi^2$ values calculated from Equation~\ref{v} using flux densities from overlapping or nearby frequencies $\nu_1$ and $\nu_2$.}
\begin{tabular}{cccc}
\hline\hline
Source & $\nu_1$ & $\nu_2$ & $\chi^2$\\
& (MHz) & (MHz) & \\
\hline
CI0008 & 153 & 151 & 0.81\\
CI0008 & 843 & 844 & 0.02\\
CI0020 & 153 & 155 & 0.11\\
CI0020 & 843 & 844 & 0.74\\
CI0020 & 9000 & 9253 & 0.00\\
s415 & 153 & 155 & 0.04\\
s895 & 153 & 155 & 0.82\\
\hline
\label{var_tab}
\end{tabular}
\end{center}
\end{table}

\subsection{Modelling the radio spectra}
\label{spectral_models}

Since CI0008 was strongly detected at low frequency, we used the 20 GLEAM sub-bands in the modelling. For all other sources, we used the GLEAM deep flux density from 200 MHz, as well as the three averaged flux density measurements (see Section~\ref{GLEAM}). We used all other flux density measurements that were available, up to 34 GHz. Since each source was selected to be unresolved at 5.5/9 GHz (i.e. $\lesssim 2\arcsec$) and since they were detected at the mas scales of the LBA observations, we did not expect any of the short-baseline flux density measurements to suffer from resolution effects. Therefore, we used all available flux density measurements. However, we discarded all measurements where the flux density uncertainty was larger than the flux density, which was the case for a few of the GLEAM deep band measurements. Where a source was undetected at low frequency, we used an upper limit on the flux density of $2\sigma$. We emphasize that only six such upper limits or averaged measurements are used from GLEAM, for sources CI0020, s150, s895 and s415, which do not significantly constrain the turnover or the general shape of the spectra, and therefore have little affect our analysis and conclusions.

The models we fit to the radio spectra of each source used the same procedure as in \citet{Tin15}, which used a non-linear least-squares fitting routine that applied the Levenberg-Marquardt algorithm. The fitting routine produces a covariance matrix, from which we took the square root of the diagonal terms as uncertainties, representing the $1\sigma$ confidence interval.

\subsubsection{Spectral models}

\label{bic}

\begin{figure*}
\begin{center}
\includegraphics[clip,trim=0mm 5mm 5mm 5mm,width=0.45\textwidth]{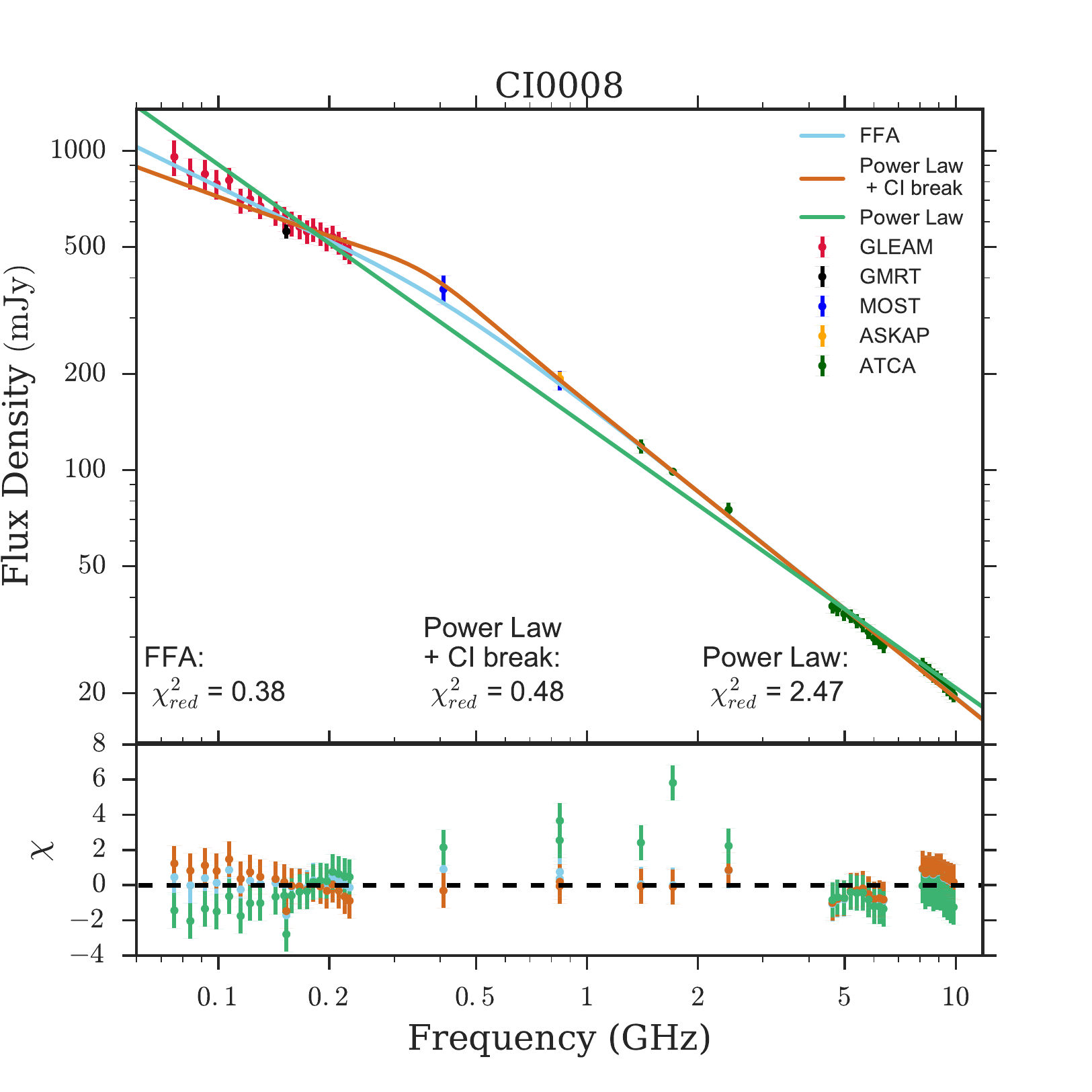}\includegraphics[clip,trim=0mm 5mm 5mm 5mm,width=0.45\textwidth]{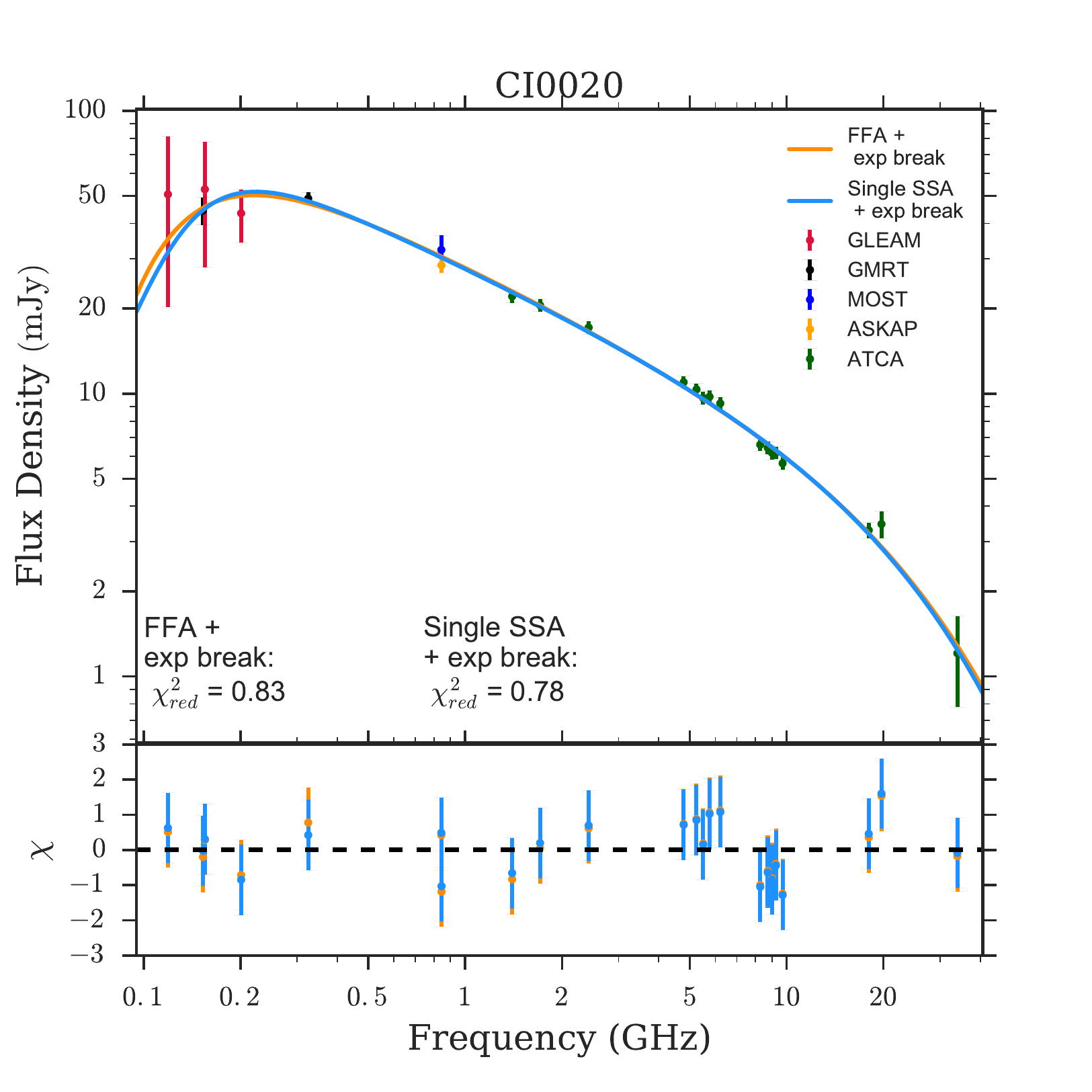}
\includegraphics[clip,trim=0mm 5mm 5mm 5mm,width=0.45\textwidth]{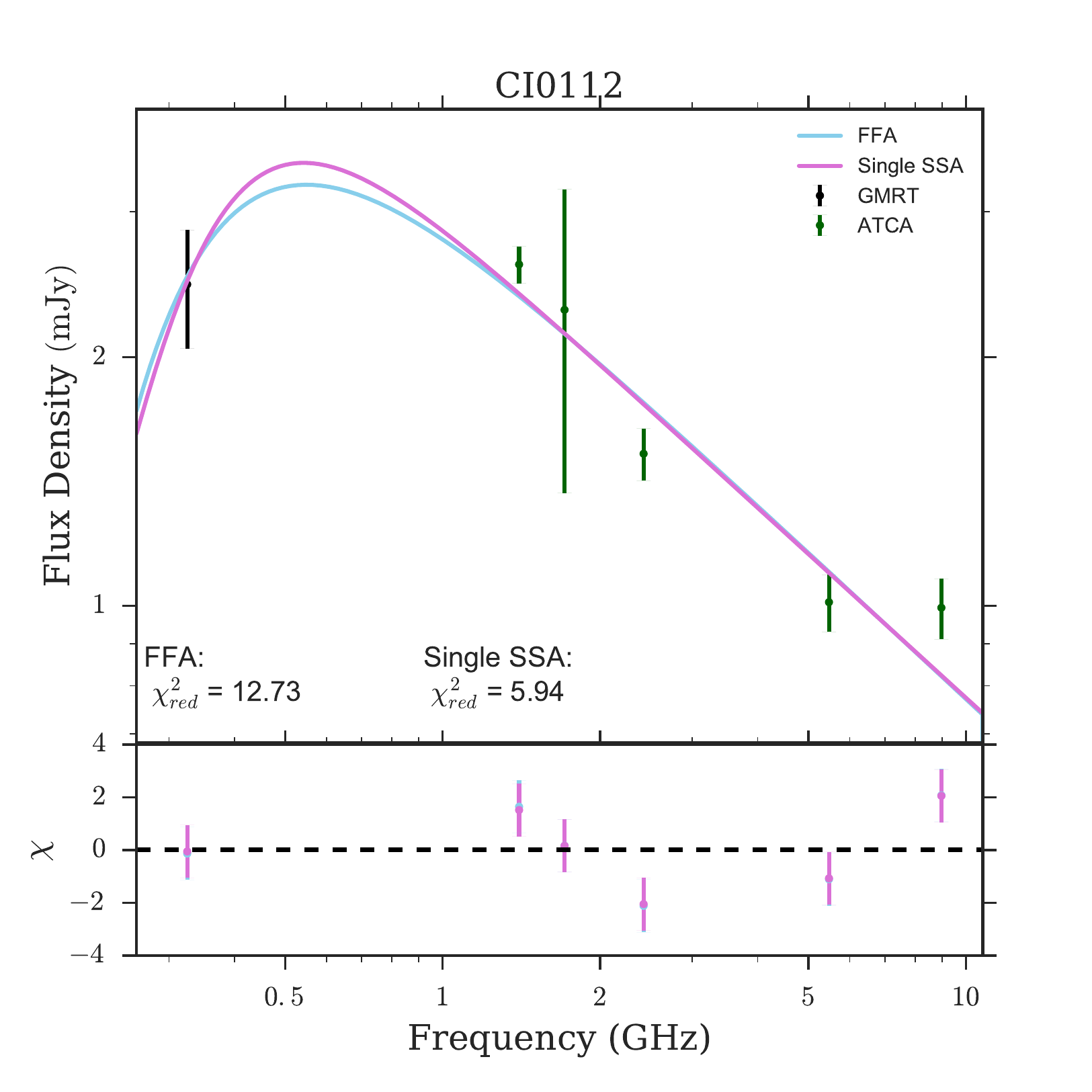}\includegraphics[clip,trim=0mm 5mm 5mm 5mm,width=0.45\textwidth]{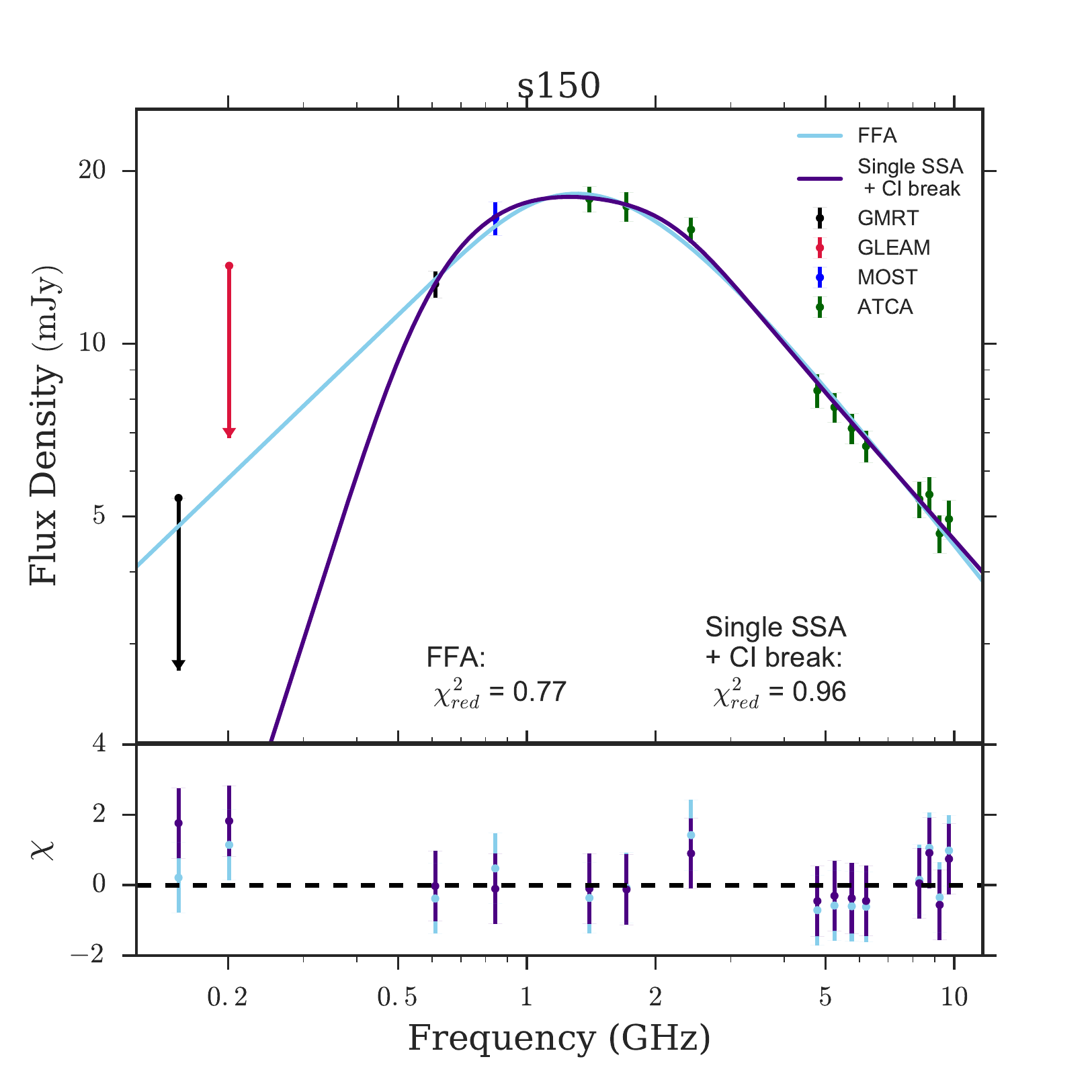}
\includegraphics[clip,trim=0mm 5mm 5mm 5mm,width=0.45\textwidth]{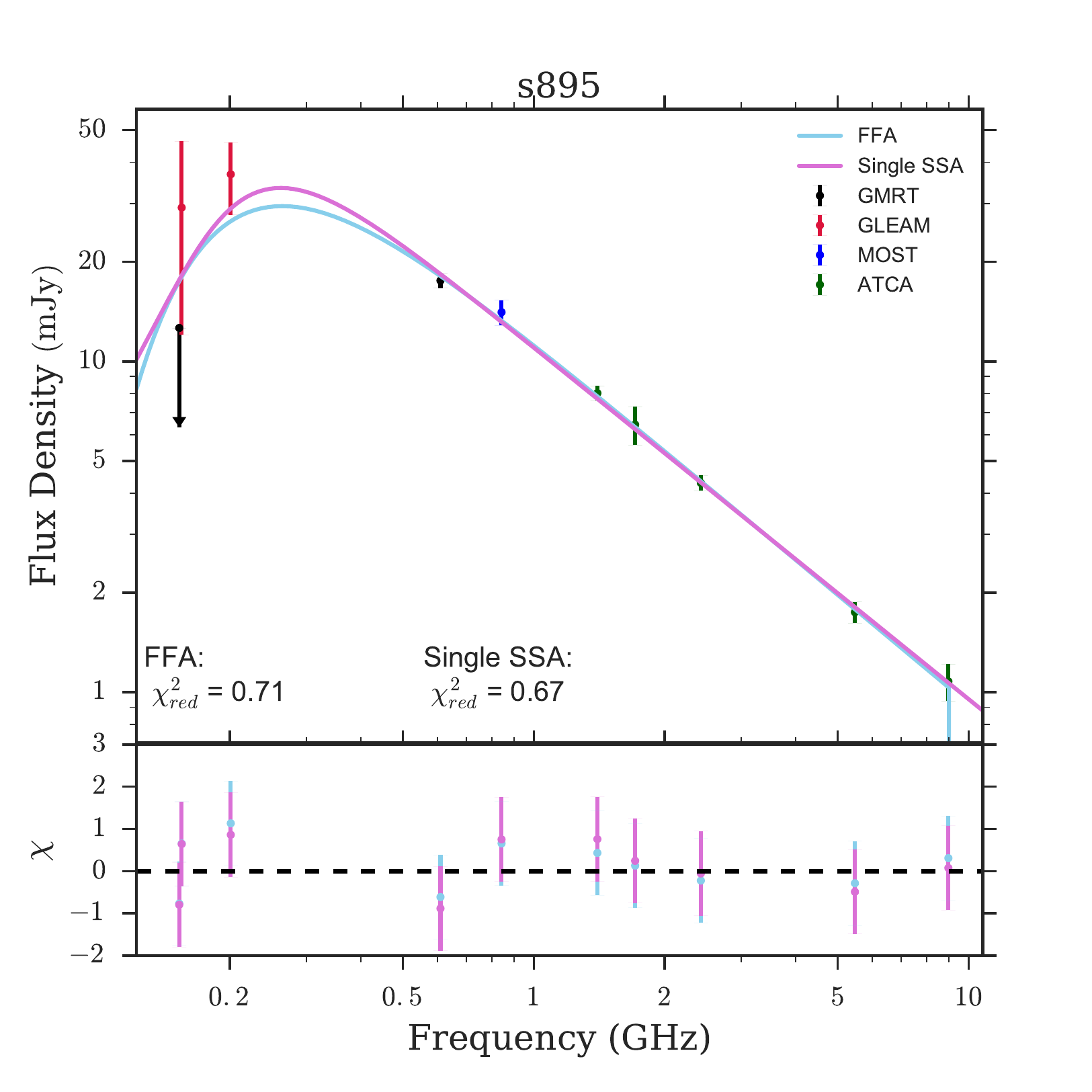}\includegraphics[clip,trim=0mm 5mm 5mm 5mm,width=0.45\textwidth]{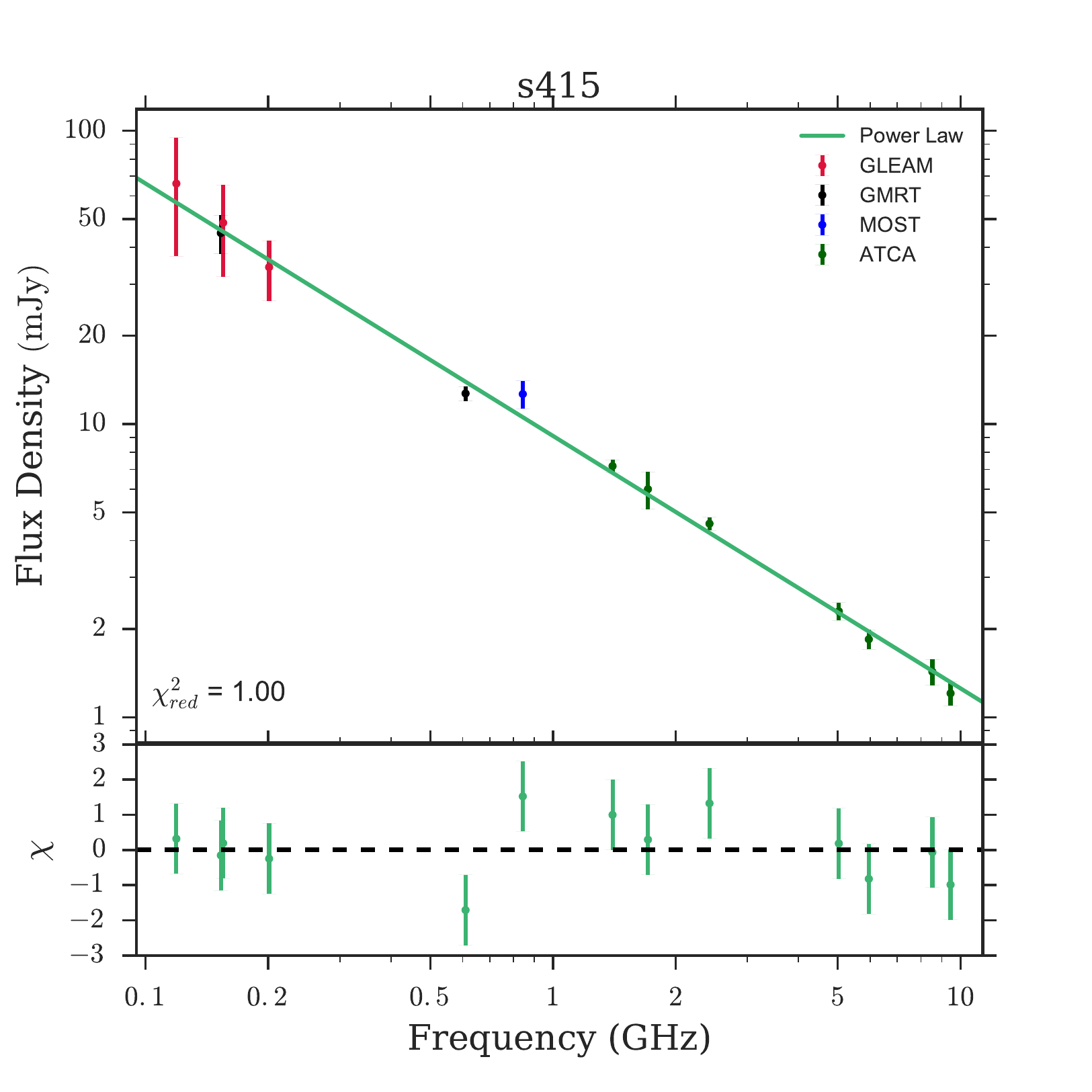}
\caption{The models fit to the radio spectrum for all sources, listed in Table~\ref{ATLAS_models_table_SSA} and \ref{ATLAS_models_table}. Upper limits at $2\sigma$ are shown by the downward arrows. A power law is fit to CI0008 for reference. We interpret CI0112 as a variable quasar. Its 1.71 GHz flux density uncertainty is large due to the way it was derived (see Section~\ref{ATLAS}).}
\label{ATLAS_models_figs}
\end{center}
\end{figure*}

Here we use the \citet{bic97} FFA model\footnote{From this point onwards, we simply refer to this model as `FFA'} given by Equation~\ref{FFA} and the homogenous SSA model given by Equation~\ref{SSA_eqn} to test whether the spectra of low-luminosity GPS and CSS sources can be represented using FFA or the widely favoured homogenous SSA model with or without spectral breaks. The models fit to our sources are shown in Fig.~\ref{ATLAS_models_figs} and summarised in Table~\ref{ATLAS_models_table_SSA} and \ref{ATLAS_models_table}. Only one FFA model requires a spectral break, while all but one of the alternative SSA and power law models do require a spectral break.

To evaluate the models, we use the Bayesian information criterion (${\rm BIC}$), calculated from the likelihood function ($\pazocal{L}$), the number of free parameters ($p$) in the given model ($f$), and the number of data points ($N$), each with a flux density ($S_{\nu_i}$) and uncertainty ($\sigma_i$) at frequency $\nu$, given by:

\begin{equation}
{\rm BIC} = -2\ln\pazocal{L} + p\ln N,
\end{equation}

\noindent where

\begin{equation}
\pazocal{L} = \prod^{N}_{i=1} \frac{1}{\sigma_i \sqrt{2\pi}}\exp{\left(-\frac{1}{2\sigma_i^2}(S_{\nu_i} - f(\nu_i))^2\right)}.
\end{equation}

\noindent When comparing two models for the same source, a $\Delta {\rm BIC} = {\rm BIC}_{\rm model 1} - {\rm BIC}_{\rm model 2} > 2$ is interpreted as positive evidence in favour of model 2, a $\Delta {\rm BIC} > 6$ as strong evidence, and a $\Delta {\rm BIC} > 10$ as very strong evidence, where the model with the lowest ${\rm BIC}$ is preferred \citep{Bayes}. 

The models were chosen based on the following decision process, each of which gave a smaller reduced $\chi^2$ value and a lower ${\rm BIC}$ value. If no curvature was seen within the spectrum, a power law was fit, otherwise both FFA and SSA were fit. If the high-frequency spectra departed from a power law, we additionally fit a spectral break. A CI or exponential break was chosen based on a combination of visual inspection of the radio spectra and the lowest reduced $\chi^2$ and ${\rm BIC}$ values. 

Source s415 did not show any curvature, so a power law was fit. Sources CI0020, s895, and s150 showed clear curvature and were fit with both SSA and FFA models. 

Source CI0020 gave $\Delta {\rm BIC}$ values that indicated very strong evidence in favour of models that included an exponential break compared to those that didn't. The FFA model including an exponential break gave $\Delta {\rm BIC} = 26.3$ compared to the FFA model without a break and $\Delta {\rm BIC} = 13.5$ compared to the FFA model with a CI break, while the SSA model including an exponential break gave $\Delta {\rm BIC} = 45.2$ compared to the SSA model and $\Delta {\rm BIC} = 17.5$ compared to the SSA model with a CI break.

In the case of s150, SSA could only be reasonably fit with a CI break included, which gave ${\rm BIC}_{\rm SSA} - {\rm BIC}_{\rm SSA + CI~break} = 10.9$, and only $\Delta {\rm BIC} = 1.0$ in the case of an exponential break. However, the location of the break frequency is strongly affected by the sampling of the spectrum, which has large gaps between measurements. For the FFA model, including a CI break gave a slightly improved $\chi^2$ value, but was not preferred according to the ${\rm BIC}$ values.

Source CI0008 showed a feature at low frequency that was either a spectral break or a very small amount of curvature, possibly close to a turnover. The spectrum was poorly fit by a power law, but gave a $\Delta {\rm BIC} = 85.6$ in favour of a power law with a CI break, which we fit to the spectrum. SSA could not account for a small amount of curvature, and therefore, we also fit an FFA model, which does not constrain a turnover, but rather a possible absorption feature caused by low density clouds.

Based on the flat spectrum given by the simultaneous 5.5 and 9.0 GHz measurements of source CI0112, we suggest it is a variable quasar (see Section~\ref{notes}), which is consistent with its highly compact VLBI emission, meaning the multi-epoch spectra we have measured does not represent its intrinsic spectrum. We fit FFA and SSA models to its spectrum, but discard it in the following analysis.

\subsection{Turnover$-$linear size relation}

Here we compare the turnovers and linear sizes of our sample to those from \cite{2014MNRAS.438..463O}, who brought together eight samples from the literature to compile a sample of young radio sources spanning a large range of linear sizes.

\begin{figure}
\begin{center}
\vspace{-4mm}
\includegraphics[width=0.5\textwidth]{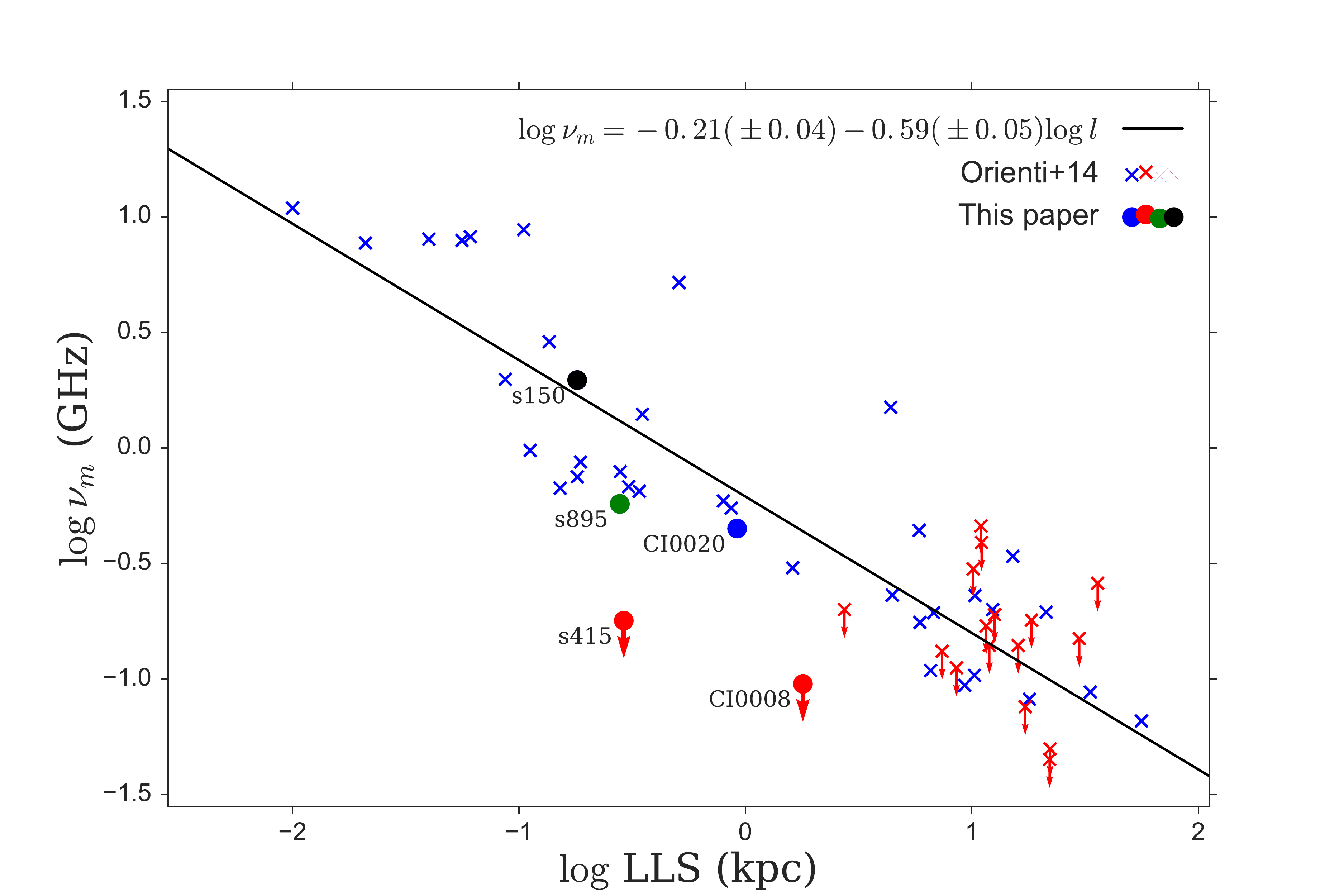}
\vspace{-5mm}
\caption{The turnover-linear size relation for our sample (circles) compared to the sample from \protect\cite[][crosses]{2014MNRAS.438..463O}. All turnover frequencies are shown in the rest frame, taken from the models in Table~\ref{ATLAS_models_table} where more than one model was fit. Sources with upper limits on the turnover frequency are shown in red and with downward arrows. The source in green (s895) has a photometric redshift, and the source in black (s150) uses an estimated redshift of $z = 0.5$. The solid line is given by Equation~\ref{linear_size_turnover}.}
\label{turnover_linear_size_VLBI}
\vspace{-5mm}
\end{center}
\end{figure}

\begin{figure}
\begin{center}
\vspace{-5mm}
\includegraphics[width=0.5\textwidth]{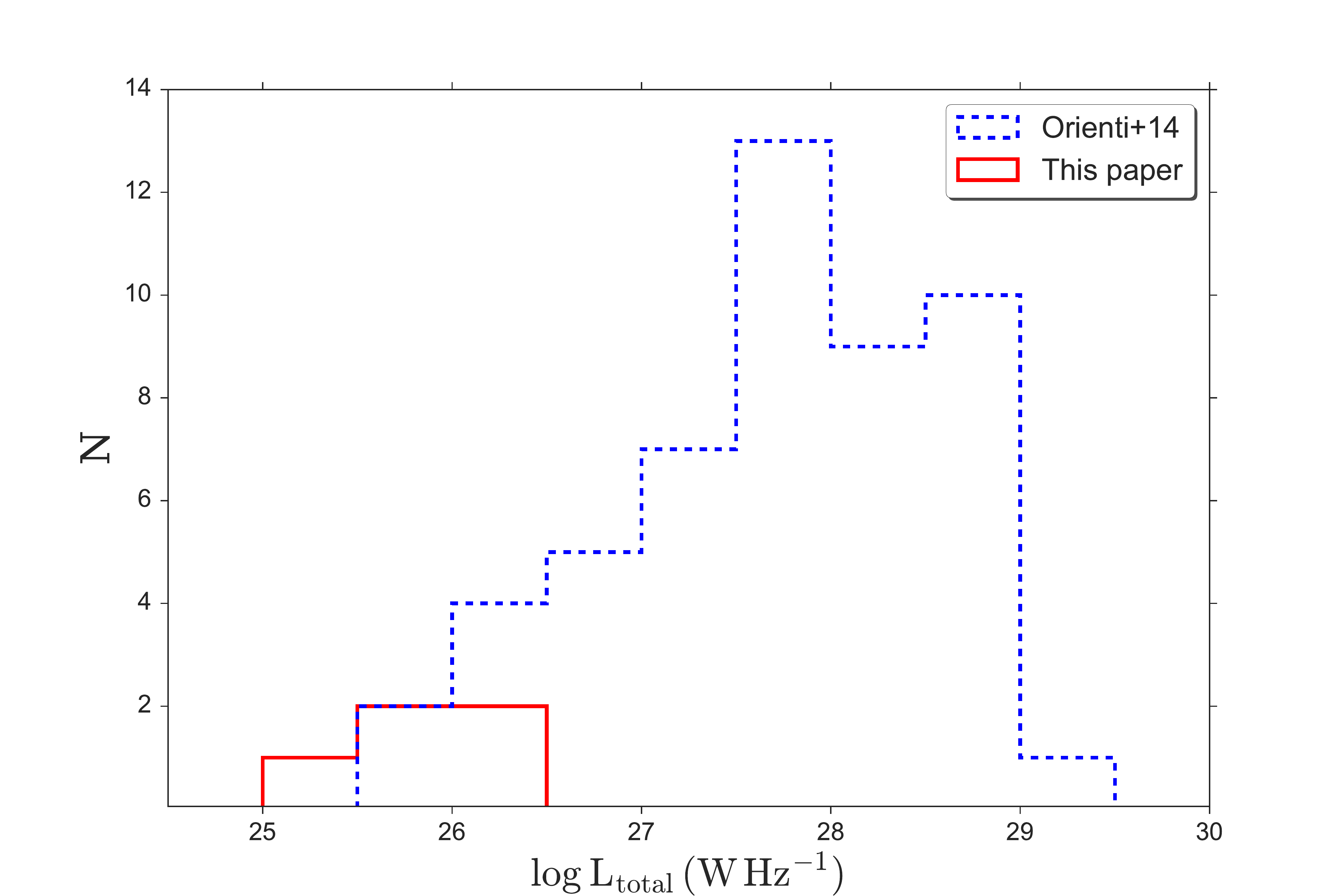}
\vspace{-5mm}
\caption{The distribution of 375 MHz total luminosities of our sample (solid red bars -- derived by extrapolating from the optically-thin spectral index) compared to the sample from \protect\cite{2014MNRAS.438..463O} (dashed blue bars), where s150 uses an estimated redshift of $z = 0.5$.}
\label{L_tot_VLBI}
\vspace{-5mm}
\end{center}
\end{figure}

Fig.~\ref{turnover_linear_size_VLBI} shows the turnover-linear size relation (see Section~\ref{turnover_linear_size}) for the sources with redshifts, using upper limits for sources that do not turn over within our data range. Fig.~\ref{L_tot_VLBI} shows the distribution of total luminosities, where the 375 MHz luminosities of our sources were derived by extrapolating from the optically-thin spectral index. This figure shows that our VLBI sources are less luminous in general than those from \cite{2014MNRAS.438..463O}, which represent the typical luminosities of the known GPS and CSS population. These figures show that low-luminosity GPS and CSS sources also follow the turnover-linear size relation. CI0008 and s415 may be considered outliers, in somewhat of a unique phase space, which may suggest they turn over at $\lesssim$100 MHz, or that the turnover frequency and linear size of low-luminosity CSS sources do not correlate in the same way as for brighter samples. 

\begin{figure}
\begin{center}
\vspace{-5mm}
\includegraphics[width=0.5\textwidth]{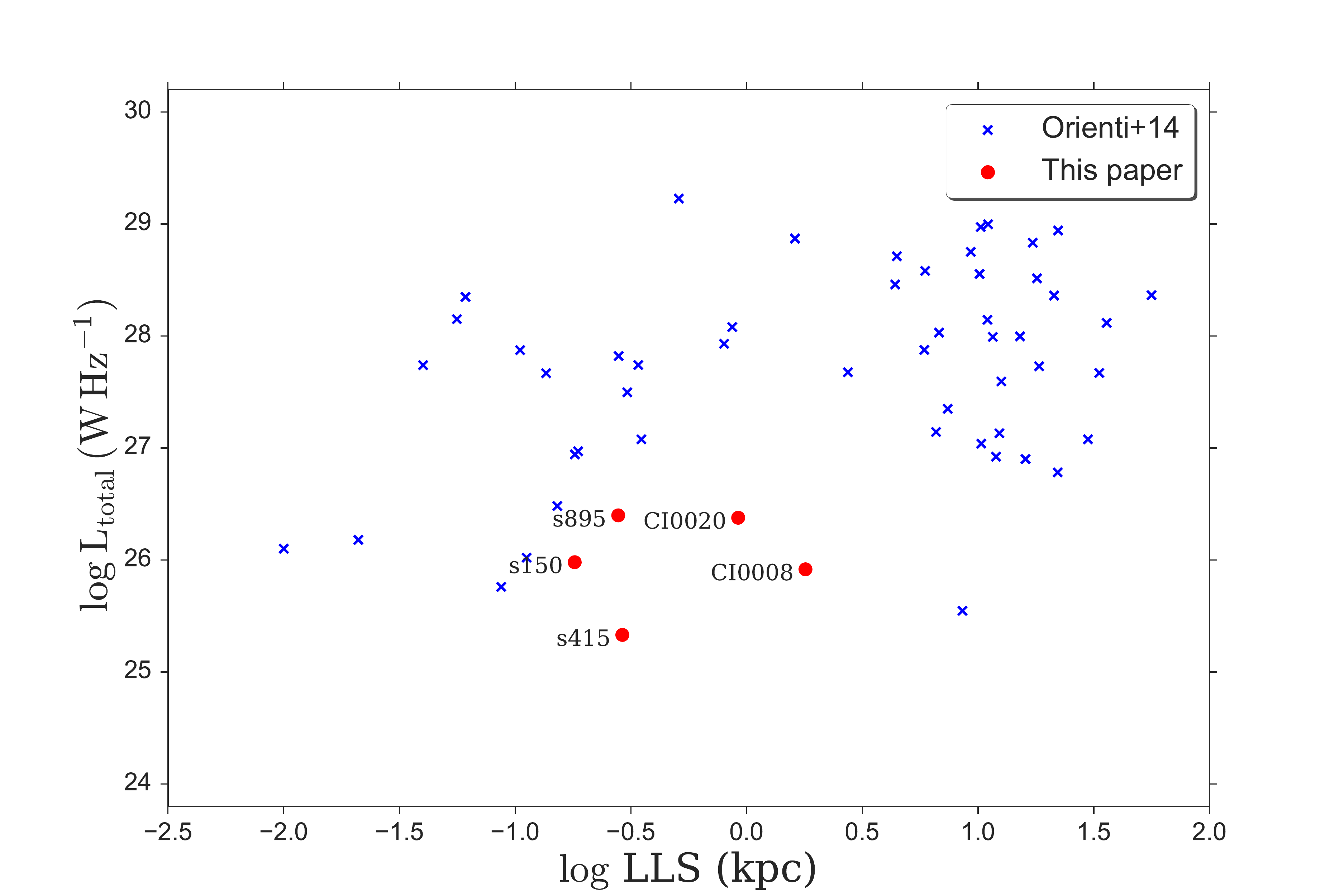}
\vspace{-5mm}
\caption{The total 375 MHz luminosities as a function of the largest linear size for our sample (red circles) compared to the sample from \citet[][blue crosses]{2014MNRAS.438..463O}, where the 375 MHz luminosities of our sources were derived by extrapolating from the optically-thin spectral index, and s150 uses an estimated redshift of $z = 0.5$.}
\label{L_tot_LLS_VLBI}
\end{center}
\end{figure}

Furthermore, the total luminosity as a function of the largest linear size (see Fig.~\ref{L_tot_LLS_VLBI}) shows that our sources are very under-represented amongst existing samples of GPS and CSS sources, being low luminosity sources in the late CSO or early MSO stage.

\cite{An12} found that the kinematic ages of their sample of CSOs followed the trend $l \propto t_{\rm kin}^{3/2}$ years, which we can express as $l = c \cdot t_{\rm kin}^{3/2}$ years, where $l$ is the linear size and $t_{\rm kin}$ is the kinematic age in the source rest frame, derived from the hotspot angular

\begin{landscape}
\begin{center}

\begin{table}
\centering
\caption{Radio spectrum SSA and power law model parameters fit to our sources. Listed is the fitted model, the synchrotron normalisation parameter $a$, the electron energy distribution $\beta$, the spectral index $\alpha$, the frequency $\nu_0$ found during the fitting (i.e. the parameter from Equation~\ref{SSA_eqn} -- see Section~\ref{sec_SSA}), the break frequency $\nu_{br}$, the turnover frequency $\nu_m$, the flux density at the spectral peak $S_{\nu_m}$, the reduced $\chi^2$ value, the degrees of freedom (DOF), and the difference in ${\rm BIC}$ values defined as ${\rm BIC}_{\rm model1} - {\rm BIC}_{\rm model2}$, where model 1 is from Table~\ref{ATLAS_models_table} and model 2 is from this table. The values are given to two significant figures, and the uncertainties are the $1\sigma$ errors, given to the same number of decimal places as the value. The individual models are shown in Fig.~\ref{ATLAS_models_figs}.}
\label{ATLAS_models_table_SSA}
\begin{tabular}{cccccccccccc}
\hline 
Source & Model & $a$ & $\beta$ & $\alpha$ & $\nu_0$ & $\nu_{br}$ & $\nu_{m}$ & $S_{\nu_m}$ & $\chi^2_{red}$ & DOF & $\Delta {\rm BIC}$\\
& & (mJy) & & & (MHz) & (GHz) & (MHz) & (mJy) & &\\
\hline
  CI0008 & Power law with CI break & 410 $\pm$ 14 & & $-$0.43 $\pm$ 0.01 & & 0.36 $\pm$ 0.02 & $<$ 76 & $>$ 810 & 0.48 & 43 & $-$0.94\\
  CI0020 & SSA with exp. break & 74 $\pm$ 3 & 2.0 $\pm$ 0.1 & & 160 $\pm$ 13 & 26 $\pm$ 4 & 220 & 52 & 0.76 & 18 & 3.3\\
  CI0112 & SSA & 5.0 $\pm$ 0.9 & 2.2 $\pm$ 0.2 & & 400 $\pm$ 77 & & 540 & 3.4 & 5.9 & 2 & 2.7\\
  s415 & Power law & 3500 $\pm$ 560 & & $-$0.86 $\pm$ 0.02 & &  &  $<$ 120 & $>$ 57 & 1.0 & 10 &\\
  s150 & SSA with CI break & 20 $\pm$ 1 & 1.3 $\pm$ 0.1 & & 730 $\pm$ 40 & 2.2 $\pm$ 0.4 & 1300 & 18 & 0.96 & 10 & $-$1.9\\
  s895 & SSA & 50 $\pm$ 4 & 3.1 $\pm$ 0.1 & & 230 $\pm$ 15 & & 260 & 33 & 0.67 & 6 & 1.8\\
\hline
\end{tabular}
\end{table}

\vspace{20mm}

\begin{table}
\centering
\caption{Radio spectrum FFA parameters fit to our sources. Listed is the fitted model, the synchrotron normalisation parameter $a$, the spectral index $\alpha$, the optical depth parameter $p$, the frequency $\nu_0$ found during the fitting (i.e. the parameters from the \citet{bic97} model $-$ see Equation~\ref{FFA} and Section~\ref{sec_FFA}), the break frequency $\nu_{br}$, the turnover frequency $\nu_m$, the flux density at the spectral peak $S_{\nu_m}$, the reduced $\chi^2$ value, and the degrees of freedom (DOF). The values are given to two significant figures, and the uncertainties are the $1\sigma$ errors, given to the same number of decimal places as the value. Where the uncertainty is greater than the fitted value, we list `-'. The individual models are shown in Fig.~\ref{ATLAS_models_figs}.}
\label{ATLAS_models_table}
\begin{tabular}{ccccccccccc}
\hline 
Source & Model & $a$ & $\alpha$ & $p$ & $\nu_0$ & $\nu_{br}$ & $\nu_{m}$ & $S_{\nu_m}$ & $\chi^2_{red}$ & DOF\\
& & (mJy) & & & (MHz) & (GHz) & (MHz) & (mJy) & &\\
\hline
  CI0008 & FFA & 380 $\pm$ 51 & $-$0.93 $\pm$ 0.01 & $-$0.83 $\pm$ 0.02 & 400  $\pm$  62 & & $<$ 76 & $>$ 900 & 0.38 & 42\\
  CI0020 & FFA with exp. break & - & $-$0.54 $\pm$ 0.03 & - & - & 28  $\pm$ 5 & 220 & 50 & 0.83 & 17\\
  CI0112 & FFA & - & $-$0.59 $\pm$ 0.20 & - & - & & 550 & 3.2 & 13 & 1\\
  s150 & FFA & 27  $\pm$ 2 & $-$0.95 $\pm$ 0.05 & $-$0.21  $\pm$  0.08 & 1500  $\pm$  190 &  &  1300 & 18 & 0.77 & 10\\
  s895 & FFA & - & $-$1.1 $\pm$ 0.0 & - & - &  &  270 & 30 & 0.71 & 5\\
\hline
\end{tabular}
\end{table}

\end{center}
\end{landscape}

\noindent separation velocity $\mu = \theta / t_{\rm kin} (1 + z)$, where $\theta$ is the angular size, which we take as the fitted major axis from the VLBI image. 

\begin{figure}
\begin{center}
\vspace{-4mm}
\includegraphics[width=0.5\textwidth]{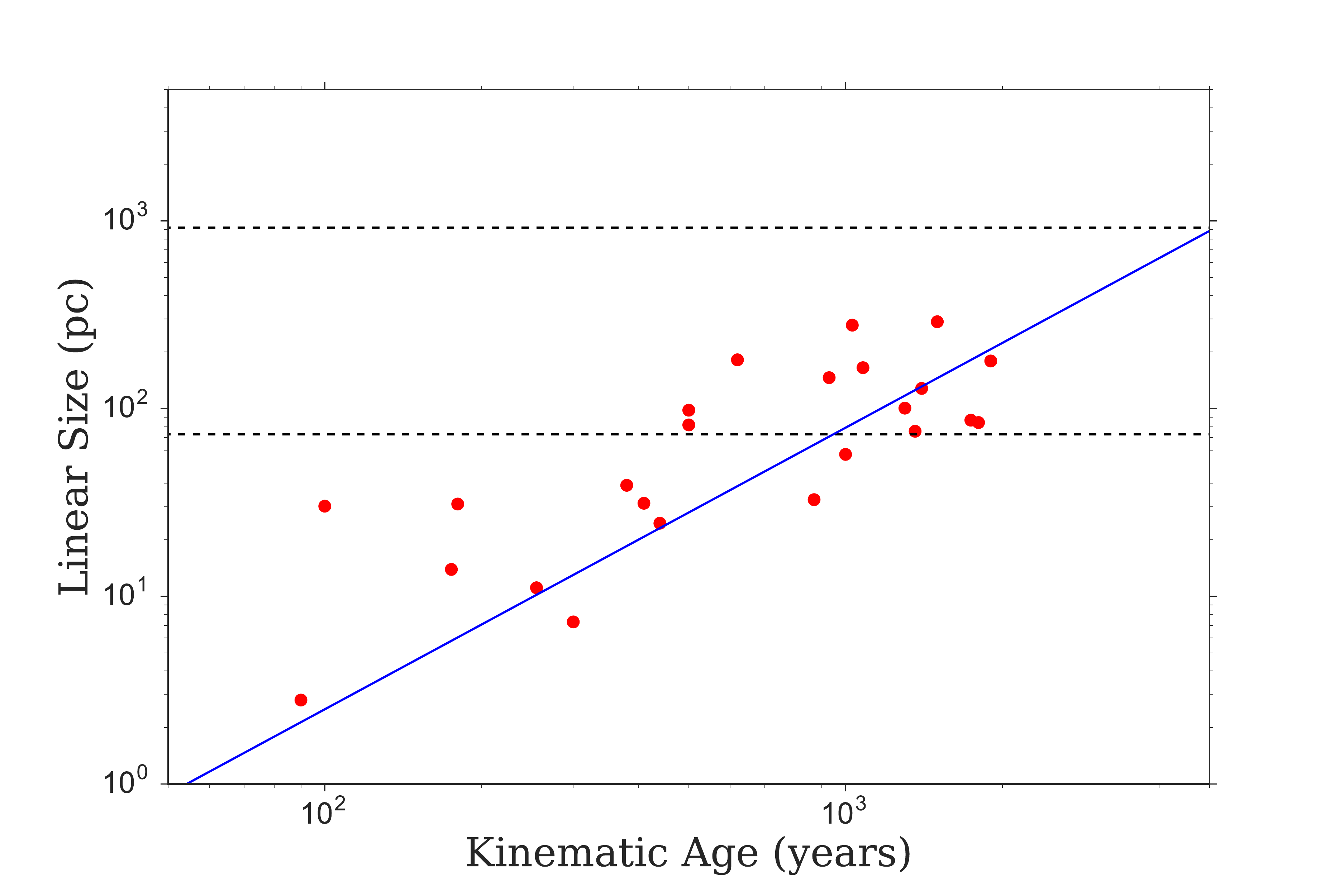}
\vspace{-5mm}
\caption{The fit to the CSOs from \protect\cite{An12}, given by Equation~\ref{scaling}, and shown by the solid blue line. We use this to derive statistical model ages for the CSOs in our sample, which range in LLS from 73 to 920 pc, as shown by the dashed black lines.}
\label{model_age_fig}
\vspace{-5mm}
\end{center}
\end{figure}

Performing a least-squares fitting routine using the kinematic ages and linear sizes from \cite{An12} gives $c = 0.0025 \pm 0.0004$, where the uncertainty is the 1$\sigma$ error taken from the diagonal of the covariance matrix, as summarised in Section~\ref{spectral_models}. Therefore, assuming our low-luminosity sources scale in the same way, we derive statistical {\it model ages} for the CSOs in our sample, given by

\begin{equation}
l = (0.0025 \pm 0.0004)~t_{\rm model}^{3/2},
\label{scaling}
\end{equation}

\noindent as shown by the fit in Fig.~\ref{model_age_fig}.

If our CSOs are significantly frustrated, the model ages will be significantly underestimated, since the jets will have been expanding at a lower rate for much longer. We may expect a similar effect if the jets are low in power or recurrent. Therefore, if frustration or recurrence plays a significant role, we can expect the model ages to be lower than the true age of the source.

\section{Discussion of individual sources}

\label{GPS_discussion}

\label{notes}

\subsection{CI0008}

CI0008 is resolved into two components with VLBI and has the appearance of a classic double, with two mini-lobes spanning $\sim$450~mas. The two components yield a hotspot emission ratio of 1.24, suggesting a slight asymmetry. At a redshift of 0.256, it has a LLS of 1.8 kpc, classifying it as an MSO, and a radio luminosity of $L_{\rm 1.4~GHz} = 2.4 \times 10^{25}$ W Hz$^{-1}$.

Its morphology is consistent with that of a small-scale edge-brightened FR II, and less consistent with a small-scale FR I, in which we would expect to detect a central core. This is consistent with predictions that the precursors of FR I galaxies are less luminous than $10^{25}$ W Hz$^{-1}$ \citep{2015MNRAS.448..252T,2015arXiv151201851S}. 

CI0008 has an intrinsic turnover frequency of 
$<$ 100 MHz, 
classifying it as a CSS source. 

As shown in Fig.~\ref{ATLAS_models_figs}, a power law across the entire spectrum produces a poor fit, since the spectrum deviates from a power law amongst the GLEAM measurements at $<$~250 MHz. This feature could indicate a spectral break or a small amount of curvature, possibly related to the turnover. If this feature is the tail end of the turnover, homogeneous SSA cannot account for it, since the curvature shown by the data would be too small. The low frequency data are reasonably fit by a power law with a CI break at 
$\sim$360 MHz, 
but with increasingly worse residuals between the model and the GLEAM measurements towards lower frequency, and a very shallow injection spectral index of $\alpha_{\rm inj} = -0.43$. If this feature is a spectral break and not part of a turnover, CI0008 must have an intrinsic turnover at much lower frequencies than 95 MHz, making it even more of an outlier on the turnover-linear size diagram than it already is in Fig.~\ref{turnover_linear_size_VLBI}. 

The FFA model produces a good fit, with a reduced $\chi^2$ value of 
0.38, and a 
$\Delta {\rm BIC}$ value of $-0.94$, 
giving no evidence in favour of either the FFA model or the power law model with a CI break. The FFA model predicts a well constrained value of 
$p = -0.83 \pm 0.02$. 
This $p$ value implies that amongst the absorbing clouds, many more with low emission measure exist compared to those with high emission measure, accounting for a small amount of curvature. 

Using the revised formula from \citet{2005AN....326..414B}, an angular separation of 450 mas, and an injection spectral index from the SSA model of $\alpha_{\rm inj} = -0.43$, we derive an equipartition magnetic field strength of $B_{\rm eq} \sim 4$ mG. Assuming this magnetic field strength, and $\nu_{\rm br} = 360$ MHz, Equation~\ref{spectral_age} gives a spectral age of 
$\sim$9.3 kyr. 
Comparatively, the FFA model with $\alpha_{\rm inj} = -0.93$ predicts $B_{\rm eq} \sim 3 $mG, and assuming that a break exists at $\nu_{\rm br} > 10$ GHz, gives a spectral age of
$\sim$2.7 kyr.

A small amount of curvature may suggest that CI0008 has an extremely broad turnover, covering a FWHM of tens of decades of frequency, compared to typical spectral widths of $\sim$1.2 decades for GPS sources \citep{1991ApJ...380...66O}, where the superposition of the radio spectrum of many absorbing components would cause a broadening of the turnover. If this curvature is an absorption feature that is not part of the turnover, another population of much more dense absorbing clouds may exist at much lower frequency, causing a second bump in the spectrum, such as those seen in \citet{2010MNRAS.405..887C}.

\subsection{CI0020}

CI0020 is resolved with VLBI and at a redshift of 1.03, has a LLS of 0.92 kpc, which is the size of a large CSO. It is fit with an intrinsic turnover frequency of 
$\sim$440 MHz, 
classifying it as a CSS source. Both the SSA and FFA models produce a good fit, with reduced $\chi^2$ values of 
0.76 and 0.83, respectively. 
A $\Delta {\rm BIC}$ value of 3.3 
gives marginal positive evidence in favour of the SSA model, since it has one fewer fitted parameters.

CI0020 is well modelled by the \cite{1973A&A....26..423J} model of an exponential spectral break, with very strong positive evidence in favour of the FFA and SSA models that include this break. This suggests that electron injection has ceased in the region dominating the radio spectrum. No CI break is observed within the spectrum, and since its optically-thin spectrum is so flat, we do not expect a CI break near the turnover. This suggests that CI0020 has ceased injection, or that it is dominated by emission in regions where the electrons have aged, such as the lobes, as opposed to where fresh electrons are injected, such as the jets and hotpots.



Using an upper limit on the angular separation given by $\Theta_{\rm maj} = 110$ mas, and an injection spectral index from the SSA model of $\alpha_{\rm inj} = -0.50$, we derive an equipartition magnetic field strength of $B_{\rm eq} \sim 8$ mG. Using this magnetic field strength and the exponential break frequency predicted by the SSA model of 
$\nu_{br} = 26$ GHz, 
we derive $t_{\rm off} \sim 0.3$ kyr, which is independent of the CI break frequency. Comparatively, the FFA model with $\alpha_{\rm inj} = -0.54$ predicts $B_{\rm eq} \sim 10$ mG, and $t_{\rm off} \sim 0.2$ kyr.

Assuming the jets of CI0020 scale in the same way as those from \citet{An12}, we derive 
$t_{\rm model} = 5.1_{-0.5}^{+0.6}$ kyr from Equation~\ref{scaling}. From this, we derive a hotspot angular separation velocity of 
$\mu = 11 \pm 1~\mu$as yr$^{-1}$. 

\subsection{CI0112}

CI0112 is resolved along one axis with VLBI and, at a redshift of 0.287, has a LLS of 73 pc. It has a very low luminosity of $L_{\rm 1.4~GHz} = 6.9 \times 10^{23}$ W Hz$^{-1}$. Its spectrum shows evidence of variability and is very poorly fit even by the best-fitting SSA model, with a reduced $\chi^2$ value of 5.9. The 5.5 and 9.0 GHz data were taken simultaneously and give a flat spectrum. Therefore we suggest that this source is a flat spectrum quasar, which may have undergone a flaring event during which it temporarily adopted a convex spectrum and was measured by ATLAS. This interpretation is consistent with its very compact size, typical of a quasar. The slight extension along one axis of the VLBI detection may suggest a core-jet morphology. If CI0112 were a GPS source, it would turn over in the GHz range according to Equation~\ref{linear_size_turnover}.

\subsection{s150}

s150 is resolved along one axis with VLBI, which gives a largest angular size of 29 mas. It is fit with an observed turnover of
$\sim 1300$ MHz, 
classifying it as a GPS source. It is very faint, with a flux density at the spectral peak of 18 mJy, and has a steep spectral index of 
$\alpha = -0.95 \pm 0.05$. 
Both the SSA and FFA models produce a good fit, with reduced $\chi^2$ values of 
0.96 and 0.77, respectively. 
A $\Delta {\rm BIC}$ value of $-1.9$
gives marginal positive evidence in favour of FFA.
The FFA model predicts a value of 
$p = -0.21 \pm 0.08$, 
implying somewhat of a dense environment.

Using an upper limit on the angular separation given by $\Theta_{\rm maj} = 29$ mas, an injection spectral index from the SSA model of $\alpha_{\rm inj} = -0.95$, and an estimated redshift of $z = 0.5$, we derive an equipartition magnetic field strength of $B_{\rm eq} \sim 12$ mG. Using this magnetic field strength and assuming the CI break of 2.2 GHz from the SSA model, we derive $t_s \sim 0.7$ kyr.
Comparatively, the FFA model with $\alpha_{\rm inj} = -0.95$ predicts $B_{\rm eq} \sim 10$ mG, and assuming a CI break at $\sim$10 GHz, this gives
$t_s \sim 0.4$ kyr.

Assuming a redshift of $z = 0.5$, s150 has a luminosity of $L_{\rm 1.4~GHz} = 1.8 \times 10^{25}$ W Hz$^{-1}$. Assuming s150 follows the same $l - t_{\rm kin}$ trend as the CSOs from \cite{An12} and has a redshift of $z = 0.5$, we derive 
$t_{\rm model} = 1.7 \pm 0.2$ kyr and $\mu = 11 \pm 1~\mu$as yr$^{-1}$. 

\subsection{s415}

s415 is resolved with VLBI and appears to be composed of two components, suggesting a symmetric double morphology. At a redshift of 0.5066, it has a LLS of 0.29 kpc, which is the size of a typical CSO. The power law produces a reasonable fit, with a reduced $\chi^2$ value of 
1.00. 
It is fit with an intrinsic turnover frequency of 
$\lesssim$ 180 MHz, 
classifying it as a CSS source. It is faint in the radio, with a flux density at the spectral peak of 
$>$ 57 mJy, and has a typical CSS spectral index of 
$\alpha = -0.86 \pm 0.02$. 
It has a relatively low luminosity of $L_{\rm 1.4~GHz} = 7.3 \times 10^{24}$ W Hz$^{-1}$.

Using an upper limit on the angular separation given by $\Theta_{\rm maj} = 47$ mas and an injection spectral index of $\alpha_{\rm inj} = -0.86$, we derive an equipartition magnetic field strength of $B_{\rm eq} \sim 7$ mG. Using this magnetic field strength and assuming a CI break of $> 10$ GHz, 
we derive $t_s \sim 0.7$ kyr.

Assuming s415 follows the same $l - t_{\rm kin}$ trend as the CSOs from \cite{An12}, we derive 
$t_{\rm model} = 2.4_{-0.2}^{+0.3}$ kyr and $\mu = 13 \pm 1~\mu$as yr$^{-1}$. 
This suggests the magnetic field strength is 
$B > 3$ mG, which gives $t_s < 2.5$ kyr.

\subsection{s895}

s895 is resolved with VLBI and, assuming its photometric redshift of 1.14 is correct, has a LLS of 0.28 kpc, which is the size of a typical CSO. It is fit with an intrinsic turnover frequency of 
$\sim$550 MHz, 
classifying it as a CSS source. It is faint in the radio, with a flux density at the spectral peak of 
29 mJy, 
and has a steep spectral index of 
$\alpha = -1.1 \pm 0.0$. 
However, being at high redshift, it is the second most luminous source in the sample, with $L_{\rm 1.4~GHz} = 5.9 \times 10^{25}$ W Hz$^{-1}$. Both the SSA and FFA models produce reasonable fits, with reduced $\chi^2$ values of 
0.67 and 0.71, respectively.
A $\Delta {\rm BIC}$ value of 1.8 
gives no evidence in favour of either model.

Using an upper limit on the angular separation given by $\Theta_{\rm maj} = 33$ mas and an injection spectral index of $\alpha_{\rm inj} = -1.1$, we derive an equipartition magnetic field strength of $B_{\rm eq} \sim 7$ mG. 

A CI break may exist at low frequency, giving an injection spectral index of 
$\alpha_{\rm inj} = -0.60$. Therefore, assuming a CI break of $< 610$ MHz and a magnetic field strength of 7 mG, we derive 
$t_s > 1.9$ kyr for both SSA and FFA.

Assuming s895 follows the same $l - t_{\rm kin}$ trend as the CSOs from \cite{An12}, we derive 
$t_{\rm model} = 2.3_{-0.2}^{+0.3}$ kyr and $\mu = 7 \pm 1~\mu$as yr$^{-1}$. 
This implies that the magnetic field strength is 
$B < 7$ mG, which gives $t_s > 2.4$ kyr.

\section{Summary and Conclusion}

\label{VLBI_conclusions}

We observed eight GPS and CSS candidates with the LBA and detected six of them, all of which were resolved, and one of which has the morphology of a classic double, like that of a small-scale FR II galaxy, consistent with the prediction that FR I precursors are less luminous than $10^{25}$ W Hz$^{-1}$ \citep{2015MNRAS.448..252T,2015arXiv151201851S}. Their redshifts range from $0.256 < z < 1.14$, giving $L_{\rm 1.4~GHz} = 10^{23-26}$ W Hz$^{-1}$ and $0.07 < {\rm LLS} < 1.8$ kpc. Our observations confirm that s150 is a GPS source turning over at 
$\sim$1.3 GHz, 
CI0112 may be a variable flat-spectrum quasar, while the other sources are CSS sources turning over between 
$\sim$76 MHz and $\sim$260 MHz. 
Our sources are lower in luminosity compared to typical GPS and CSS sources, with $L_{\rm 1.4 GHz} < 10^{27}$ W Hz$^{-1}$, and follow the turnover-linear size relation, showing that low-luminosity GPS and CSS sources also co-evolve in size and turnover.

We derive statistical model ages for CI0020, s150, s895 and s415, based on the \cite{An12} scaling relation of $l \propto t_{\rm kin}^{3/2}$, in the range 
$1.7 < t_{\rm model} < 5.1$ kyr. 
Based on equipartition  magnetic field strengths in the range of $4 < B_{\rm eq} < 12$ mG, we derive spectral ages in the range 
$0.7 < t_s < 2.7$ kyr, and 9.3 kyr for the broken power law model for CI0008. 
We find that CI0020 is well modelled by the \cite{1973A&A....26..423J} model of an exponential spectral break, with a turnoff time of 
$t_{\rm off} \sim 0.3$ kyr, suggesting the region dominating the radio spectrum has ceased injecting new electrons. The estimated ages of our sources are similar to those of brighter samples, suggesting that low-luminosity GPS and CSS sources are young and evolving. 

The GPS and CSS sources are well fit by the homogeneous SSA model, as well as the \citet{bic97} inhomogeneous FFA model. A $\Delta {\rm BIC}$ of 3.3 and $-$1.9 gives marginal positive evidence in favour of an SSA model for source CI0020, and an FFA model for source s150, respectively. but no evidence in favour of an FFA or SSA model in all other cases. All but one of the FFA models do not require a spectral break to account for the radio spectrum, while all but one of the alternative SSA and power law models do require a spectral break to account for the radio spectrum.

CI0008 deviates from a power law at low frequency, which cannot be accounted for by a small amount of curvature from SSA, suggesting that either FFA or electron ageing is producing this feature, which may be the beginning of the optically-thick region of a very broad, very shallow turnover, or a low-frequency spectral break. 

We conclude that our faint and low-luminosity population of GPS and CSS sources are similar to brighter samples in terms of their spectral shape, turnover frequencies, linear sizes and ages. However, we could not test for a morphological difference except for CI0008, which appears as a scaled-down FR~II, consistent with predictions \citep{2015MNRAS.448..252T,2015arXiv151201851S}.

For this particular class of AGN, the Evolutionary Map of the Universe \cite[EMU;][]{EMU} and other ASKAP, MWA and MeerKAT surveys will revolutionise our understanding. MWA observations will enable the turnovers and optically-thick spectra to be thoroughly measured, enabling the different spectral models be tested over large samples. Early science for ASKAP will give continuous radio coverage from $\sim700-1800$ MHz, enabling GPS and CSS samples to be selected from within one set of observations. EMU will detect several million GPS and CSS sources, and uncover the low-luminosity population. Studying these samples of AGN will enable their evolution to be characterised across a vast range of cosmic time and evolutionary ages.

\section*{Acknowledgements}

The authors would like to thank Bjorn Emonts and Alexandre Beelen for permission to use the 34 GHz ATCA observations of the ECDFS, and the referee for their helpful comments that improved this paper.

The Australia Telescope Compact Array and Long Baseline Array are part of the Australia Telescope National Facility, which is funded by the Australian Government for operation as a National Facility managed by CSIRO.

The Australian SKA Pathfinder is part of the Australia Telescope National Facility, which is managed by CSIRO. Operation of ASKAP is funded by the Australian Government with support from the National Collaborative Research Infrastructure Strategy. ASKAP uses the resources of the Pawsey Supercomputing Centre. This scientific work makes use of the Murchison Radio-astronomy Observatory, operated by CSIRO. Support for the operation of the MWA is provided by the Australian Government (NCRIS), under a contract to Curtin University administered by Astronomy Australia Limited. Establishment of ASKAP, the Murchison Radio-astronomy Observatory and the Pawsey Supercomputing Centre are initiatives of the Australian Government, with support from the Government of Western Australia and the Science and Industry Endowment Fund. We acknowledge the Wajarri Yamatji people as the traditional owners of the Observatory site.

This work made use of the Swinburne University of Technology software correlator, developed as part of the Australian Major National Research Facilities Programme.

We thank the staff of the GMRT who have made these observations possible. GMRT is run by the National Centre for Radio Astrophysics
  of the Tata Institute of Fundamental Research.
  
The MOST is operated by the University of Sydney with support from the Australian Research Council and the Science Foundation for Physics within the University of Sydney.

Part of this work has been conducted with the financial support from the Australian Research Council Centre of Excellence for All-sky Astrophysics (CAASTRO), through project number CE110001020.

This research has made use of the VizieR catalog access tool, CDS, Strasbourg, France. Topcat \citep{2005ASPC..347...29T}, SAOImage DS9, NASA's Astrophysics Data System bibliographic services, and Astropy, a community-developed core Python package for Astronomy \citep{2013A&A...558A..33A}, were also used in this study.

\bibliographystyle{mnras}
\bibliography{bib}

\label{lastpage}
\end{document}